\documentclass[pra,aps,showpacs,reprint,superscriptaddress]{revtex4-1}
\usepackage{amsmath,amssymb}
\usepackage[dvipdfmx]{graphicx}
\usepackage[final]{hyperref}
\usepackage{amsmath}
\usepackage{bm}
\usepackage{braket}
\usepackage{float}
\usepackage[caption=false]{subfig}
\usepackage{mathtools}
\usepackage{siunitx}
\usepackage{lineno}

\hypersetup{
	colorlinks=true,       
	linkcolor=blue,          
	citecolor=blue,        
	filecolor=magenta,      
	urlcolor=blue   
}

\begin{document}

	
\title{Fast amplification and rephasing of entangled cat states 
\\in a qubit-oscillator system
}
\author{Z. Xiao}
\email[xiaozhihao@hotmail.com]{}
\affiliation{Hearne Institute for Theoretical Physics, and Department of Physics $\&$ Astronomy, Louisiana State
	University, Baton Rouge, Louisiana 70803, USA}
\affiliation{National Institute of Information and Communications Technology, 4-2-1, Nukui-Kitamachi, Koganei, Tokyo 184-8795, Japan}

\author{T. Fuse}
\email[tfuse@nict.go.jp]{}
\affiliation{National Institute of Information and Communications Technology, 4-2-1, Nukui-Kitamachi, Koganei, Tokyo 184-8795, Japan}

\author{S. Ashhab}
\affiliation{Qatar Environment and
	Energy Research Institute, Hamad Bin Khalifa University, Qatar Foundation, Doha, Qatar}

\author{F. Yoshihara}

\author{K. Semba}

\author{M. Sasaki}

\author{M. Takeoka}
\affiliation{National Institute of Information and Communications Technology, 4-2-1, Nukui-Kitamachi, Koganei, Tokyo 184-8795, Japan}

\author{J. P. Dowling}
\affiliation{Hearne Institute for Theoretical Physics, and Department of Physics $\&$ Astronomy, Louisiana State
	University, Baton Rouge, Louisiana 70803, USA}
\affiliation{National Institute of Information and Communications Technology, 4-2-1, Nukui-Kitamachi, Koganei, Tokyo 184-8795, Japan}
\affiliation{NYU-ECNU Institute of Physics at NYU Shanghai, Shanghai 200062, China}
\affiliation{CAS-Alibaba Quantum Computing Laboratory, USTC, Shanghai 201315, China}

\begin{abstract}
We study a qubit-oscillator system, with a time-dependent coupling coefficient, and present a fast scheme for generating entangled Schr\"odinger-cat states with large mean photon numbers
and also a scheme that protects the cat states against dephasing caused by the nonlinearity in the system. 
We focus on the case where the qubit frequency is small compared to the oscillator frequency.
We first present the exact quantum state evolution in the limit of infinitesimal qubit frequency.
We then analyze the first-order effect of the nonzero qubit frequency.
Our scheme works for a wide range of coupling strength values, including the recently achieved deep-strong-coupling regime.
\end{abstract}

\maketitle 

\section{Introduction}
The interaction of a two-level atom (qubit) with a quantized field (oscillator) has been widely studied over the past few decades.
There have been numerous experimental realizations of such systems, including superconducting circuits 
\cite{chiorescu2004coherent, wallraff2004strong, devoret2007circuit, niemczyk2010circuit, forn2010observation, yoshihara2017superconducting, yoshihara2017characteristic, didier2015fast, yin2012dynamic, Nori2017review}, and 
systems of atoms coupled to superconducting microcavities \cite{Haroche2008reconstruction, Haroche}.
Mathematically, such qubit-oscillator systems are described by the quantum Rabi model. Outside the regime where the rotating-wave approximation  (RWA) can be used, previous studies have mainly focused on systems with time-independent coupling coefficients~\cite{Irish2007gRWA, Braak2011, ashhab2010qubit, casanova2010deep}.

In this paper, we examine the qubit-oscillator system with a time-dependent coupling coefficient \cite{Schuetz2017, wang2009coupling, Clerk2018}, where the qubit frequency is small compared to that of the oscillator and the RWA is not applicable. 
We solve the dynamics of the system in the case of a general time-dependent coupling and use this solution to demonstrate schemes for generating large Schr\"odinger cat states, and for protecting them from dephasing.
Very large-size Schr\"odinger cat states are useful for quantum information processing~\cite{hofheinz2009synthesizing, vlastakis2013deterministically} and quantum enhanced sensing ~\cite{Haroche2016sensing}, for instance.
Our amplification scheme 
offers a potentially simple and fast alternative compared to previous methods \cite{hofheinz2009synthesizing, vlastakis2013deterministically}. 
A simple estimation indicates three orders of magnitude speedup in generating a 100-photon cat state, compared to Ref.~\cite{vlastakis2013deterministically}. See also Appendix C for details.
In the following, we start by presenting the state evolution in the exact analytical form in the limit of an infinitesimal qubit frequency, 
and then we examine the first-order effect of a small nonzero qubit frequency.

\section{State evolution under time-dependent coupling with infinitesimal qubit frequency}
The Hamiltonian of the combined system of the qubit and the oscillator is
\begin{equation} \begin{split}\label{eqn:Hamiltonian_full}
\hat{H}(t)=-\frac{\hbar}{2}\Delta\hat{\sigma}_z+\hbar\omega\left(\hat{a}^\dagger\hat{a}+\frac{1}{2}\right)+\hbar g(t)\hat{\sigma}_x(\hat{a}^\dagger+\hat{a}),
\end{split}\end{equation}
where $\omega$ and $\Delta$ are the frequencies of the oscillator and the qubit, respectively,
and $g(t)$ is the time-dependent coupling constant, $\hat{a}$ and $\hat{a}^\dagger$ are, respectively,
the annihilation and creation operators of the oscillator, and $\hat{\sigma}_{x, z}$ are the Pauli operators of the qubit 
\cite{RabiModel}. 
We focus on the situation where $\Delta$ is small compared to $\omega$, without assuming any condition on the coupling $g(t)$.
We will first examine the zeroth order effect of the small $\Delta$ by taking the limit of $\Delta/\omega\rightarrow0$, which 
gives
\begin{equation} \begin{split}\label{eqn:Hamiltonian_Delta_and_epsilon_dropped}
\hat{H}^{(0)}(t)=\hbar\omega\left(\hat{a}^\dagger\hat{a}+\frac{1}{2}\right)+\hbar g(t)\hat{\sigma}_x(\hat{a}^\dagger+\hat{a}).
\end{split}\end{equation}
The initial eigenstates of the Hamiltonian 
at $t=0$ are the entangled states \cite{ashhab2010qubit, casanova2010deep}:
\begin{equation} \begin{split}\label{eqn:energy_eigenstates}
\ket{E^{(0)}_{N\pm}(0)}=&\ket{+}_x\hat{D}\left({-\frac{g(0)}{\omega}}\right)\ket{N}\\
&\pm\ket{-}_x\hat{D}\left(+{\frac{g(0)}{\omega}}\right)\ket{N},
\end{split}\end{equation}
ignoring a factor of $1/\sqrt{2}$, and where $\ket{\pm}_x$ are the two qubit eigenstates of the Pauli matrix $\hat{\sigma}_x$ with eigenvalues $\pm1$, $\ket{N}$ is the $N$-photon Fock state in the oscillator, and $\hat{D}\left({\pm\frac{g(0)}{\omega}}\right)$ are displacement operators. 
Note that, in the limit of $\Delta/\omega\rightarrow0$, the energy eigenstates $\ket{E^{(0)}_{N+}(0)}$ and $\ket{E^{(0)}_{N-}(0)}$ are degenerate, with the energy eigenvalues of $E_{N\pm}^{(0)}(0)=
\hbar\omega \left(N+1/2-g^2(0)/\omega^2\right)$.

We now consider the state evolution under an arbitrary time-dependent coefficient $g(t)$. The energy eigenstates $\ket{E_{N \pm}^{(0)}(t)}$, determined by the instantaneous value of $g(t)$, do not reflect the evolution of quantum states. In other
words, an initial state $\ket{E_{N \pm}^{(0)}(0)}$ generally does not, for a time-dependent $g(t)$, evolve into 
$e^{i\phi_{N\pm}(t)}\ket{E_{N \pm}^{(0)}(t)}$, with a phase factor $\phi_{N\pm}(t)$, 
at a later time $t$. 
On the other hand, a quantum state in the form of $\ket{\tilde{E}_{N \pm}^{(0)}(0)}$ evolves into the quantum state 
$e^{i\phi_N(t)}\ket{\tilde{E}_{N \pm}^{(0)}(t)}$, where
\begin{equation} \begin{split}\label{eqn:tilde_E_of_t}
\ket{\tilde{E}^{(0)}_{N\pm}(t)}=&\ket{+}_x\hat{D}\left({-\frac{\tilde{g}(t)}{\omega}}\right)\ket{N}\\
&\pm\ket{-}_x\hat{D}\left(+{\frac{\tilde{g}(t)}{\omega}}\right)\ket{N}, 
\end{split}\end{equation}
and the phase factor $\phi_N(t)$ is given in Appendix A.
Here the complex variable $\tilde{g}(t)$ obeys the equation
\begin{equation}\label{eqn:differential_g_1} \begin{split}
\dot{\tilde{g}}(t)=i \omega (g(t)-\tilde{g}(t)). 
\end{split}\end{equation}
Note that the initial $\tilde{g}(0)$ can be set to any value. The proof is given in Appendix A. We refer to $\ket{\tilde{E}^{(0)}_{N\pm}(t)}$ as the dynamical evolution eigenstates, which take a similar form to the energy eigenstates in Eq.~(\ref{eqn:energy_eigenstates}), but with $g(t)$ replaced with $\tilde{g}(t)$. 
Now the dynamics of the system is governed by Eq.~(\ref{eqn:differential_g_1}), which shows that $\tilde{g}(t)$ does not respond instantly to changes in $g(t)$. 
As the energy eigenstates provide a convenient basis such that any initial state expressed as a superposition of eigenstates accumulates a phase factor but is otherwise unchanged under a time-independent Hamiltonian, the dynamical evolution states provide such a basis that quantum evolution can be described conveniently under time-dependent Hamiltonian. 
Note that, apart from a few special cases, the dynamical evolution eigenstates are generally not energy eigenstates.

To understand the evolution of quantum states, 
we give the following three scenarios, in which we set the initial 
$\tilde{g}(0)=g(0)$ so that $\ket{\tilde{E}_{N\pm}^{(0)}(0)}=\ket{E_{N\pm}^{(0)}(0)}$: 

\begin{figure}
\includegraphics[width=0.45\textwidth]{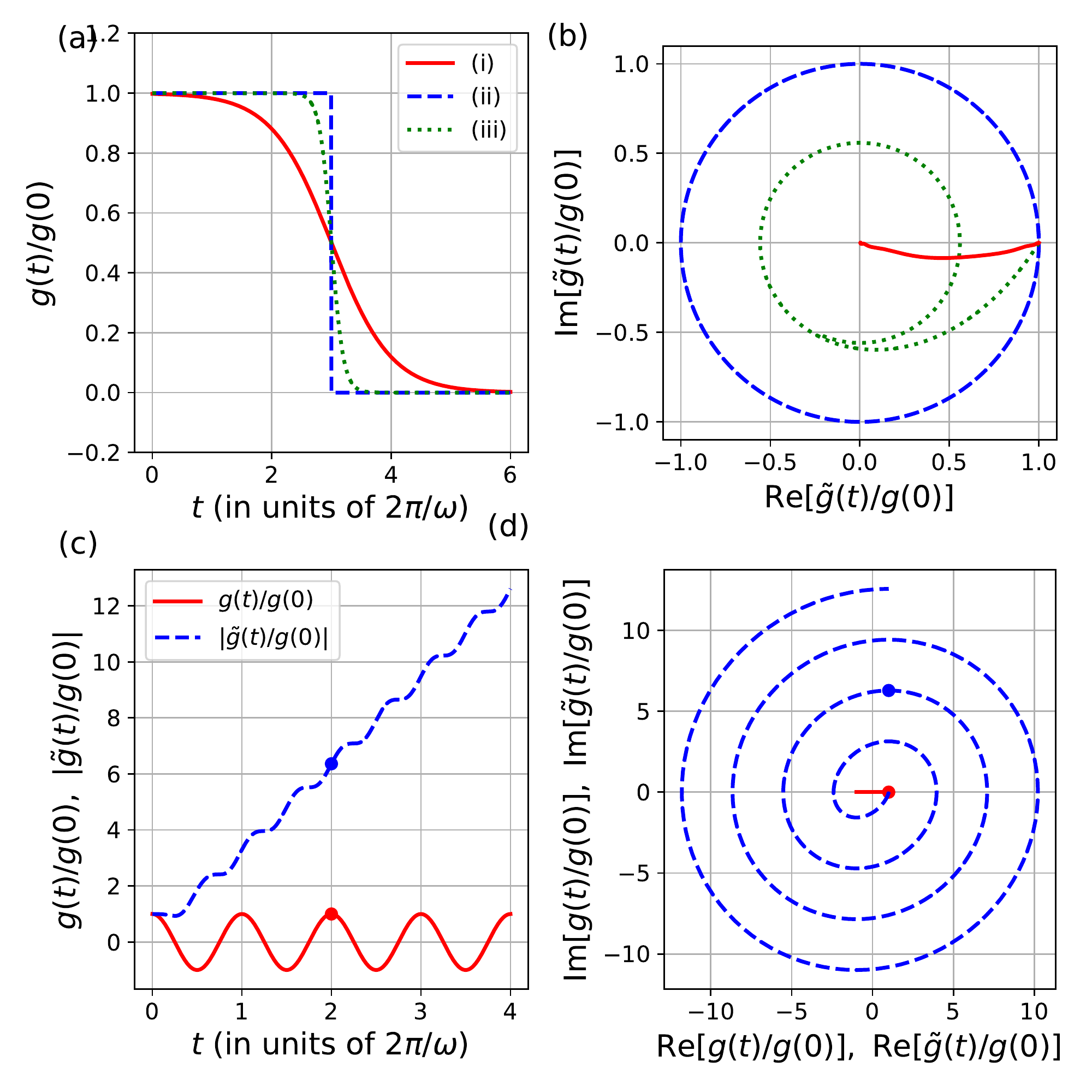}
\caption{\label{fig:gAndgTilde}
(color online) 
Evolution of $\tilde{g}(t)$ under time-dependent $g(t)$ with infinitesimal $\Delta$.
(a) Three different time dependencies of $g(t)$ as a function of time $t$.
(b) The trajectories of $\tilde{g}(t)$ corresponding to (a),
which, as seen in Eq.~(\ref{eqn:tilde_E_of_t}), determine the evolution of the dynamical evolution eigenstates.
We set the initial condition as $\tilde{g}(0)/g(0)=1$. 
(c) The time dependencies of $g(t)$ and $|\tilde{g}(t)|$ as functions of time $t$, in the case of sinusoidal driving force 
$g(t) = g(0)\cos \omega t$.
(d) The trajectories of $g(t)$ and $\tilde{g}(t)$ corresponding to (c).
The amplitude of $\tilde{g}(t)$ keeps increasing, 
showing cat-state amplification.
The  blue and red dots in (c) and (d) indicate where the modulation stops ($t=4\pi/\omega$) in the scheme shown in Fig.~\ref{fig:CatGrowthAndPreservation}.
	}
\end{figure}

(i) Suppose $g(t)$ is adiabatically changed over time [red, solid curve in Fig.~\ref{fig:gAndgTilde}(a)]. 
Any energy eigenstate 
has ample time adjust to the adiabatically changing $\hat{H}^{(0)}(t)$ and also remains an energy eigenstate
[red, solid curve in Fig.~\ref{fig:gAndgTilde}(b)].

(ii) Suppose $g(t)$ is constant at a certain value at $t<t_0$, and 
then set to zero instantaneously at $t=t_0$ [blue, dashed curve in Fig.~\ref{fig:gAndgTilde}(a)].
Since neither $\omega$ nor $\Delta$ is infinitely large, the states $\ket{\tilde{E}^{(0)}_{N\pm}(t)}$ cannot adjust instantaneously, and they remain the same at $t=t_{0+}$. 
However, $\ket{\tilde{E}^{(0)}_{N\pm}(t_{0+})}$ are no longer the energy eigenstates, 
and the states begin to evolve.
Taking the ground dynamical evolution eigenstate
\begin{equation} \begin{split}\label{eqn:tilde_E_of_t_GroundState}
\ket{\tilde{E}^{(0)}_{0\pm}(t)}=\ket{+}_x\ket{-\tilde{g}(t)/\omega}\pm\ket{-}_x\ket{+\tilde{g}(t)/\omega}
\end{split}\end{equation} 
as an example, the amplitude of the coherent state component of the state, $\ket{{\pm\frac{\tilde{g}(t_0)}{\omega}}}$ before the adjustment, should begin to revolve around the origin after the adjustment, consistent with the evolution of a regular coherent state in a free oscillator [blue, dashed circle in Fig.~\ref{fig:gAndgTilde}(b)].
When $g(t)$ is instantaneously set to a nonzero value,
$\tilde{g}(t)$ revolves around this value in the complex plane.

(iii) Now we consider the intermediate scenario. 
We assume that $g(t)$ is adjusted over a finite period of time to zero and then kept stabilized as shown in the green, dotted curve in Fig.~\ref{fig:gAndgTilde}(a).
In this scenario, $\tilde{g}(t)$ will start changing as $g(t)$ starts changing.
Its trajectory is less intuitive than the extreme scenarios, but can be understood from Eq.~(\ref{eqn:differential_g_1}).
After $g(t)$ becomes constant again, $\tilde{g}(t)$ evolves in circular motion around the new constant $g$ [green, dotted curve in Fig.~\ref{fig:gAndgTilde}(b)].

By modulating $g$ periodically on resonance with the frequency $\omega$, 
we can amplify the absolute value of the amplitude of the entangled-cat-state components of Eq.~(\ref{eqn:tilde_E_of_t_GroundState}). 
As a specific example,
the case of a sinusoidal modulation 
$g(t) = g(0)\cos \omega t$ 
is shown in Figs.~\ref{fig:gAndgTilde}(c) and \ref{fig:gAndgTilde}(d).
The magnitude $|\tilde{g}(t)|$ 
will grow linearly with time.
This behavior is easy to understand from Eq.~(\ref{eqn:differential_g_1}), whose solution corresponds to a simple harmonic oscillator being driven by an external force.
Since the photon number in the coherent state is proportional to $|\tilde{g}/\omega|^2$, it will grow quadratically as a function of time.
In this case, modulating $g(t)$ for 
two oscillator periods increases the absolute amplitude of the coherent state component by 
a factor of 6.4. 

\section{First-order effect of finite qubit frequency}
So far, we have ignored the effect of the small $\Delta$ by taking the limit of $\Delta/\omega\rightarrow0$. We now examine the first-order effect in $\Delta$ in the full Hamiltonian in Eq.~(\ref{eqn:Hamiltonian_full}). 
Note that, in recent experiments \cite{yoshihara2017superconducting, yoshihara2017characteristic}, $\Delta/\omega\approx0.1$. 
At any time $t$, we can express a general state of interest as a superposition of dynamical evolution eigenstates: $\ket{\varphi(t)}=\sum_{N,\pm}^{}C_{N\pm}(t)e^{-i N\omega t}\ket{\tilde{E}^{(0)}_{N\pm}(t)}$. 
Under the full Hamiltonian $\hat{H}(t)$, the $C_{N\pm}(t)$ generally change over time.
If we consider up to the first order in $\Delta/\omega$, 
$\bold{C}(t)\equiv
\begin{bmatrix}
C_{0+}(t)       & C_{1+}(t)  &   \dots & C_{0-}(t)  & C_{1-}(t)  &  \dots 
\end{bmatrix}
$
can be expressed as
$\bold{C}(t)=\bold{C}(0)\exp \left \{ \frac{i}{2} \int_{0}^{t}\Delta(t)[M_{\sigma_z}(t)] dt \right \}$,
where $M_{\sigma_z}(t)$ is the matrix of the operator $\hat{\sigma}_z$ in the basis 
$e^{-iN\omega t}\ket{\tilde{E}^{(0)}_{N\pm}(t)}$ at time $t$. 
See Appendix B for the derivation. 
For generality, we have made the parameter $\Delta$ time dependent [$\Delta=\Delta(t)$].

An intuitive way to understand the effect of nonzero $\Delta$ is the following. If $\Delta$ were to be considered infinitesimal, the quantized oscillator has equally spaced energy levels. A nonzero $\Delta$ disrupts such equally spaced energy levels, causing any general quantum state to dephase.
This type of dephasing is well known for coherent state solutions to the harmonic oscillator~\cite{Schrodinger1926}, 
and has been seen in experiments with single-electron Rydberg atoms~\cite{Kodach:10}.

We have shown
above
that a nonzero $\Delta$ leads to the change in $C_{N\pm}(t)$ and the dephasing.
To minimize this dephasing, we need to make $\int_{0}^{t}\Delta(t)[M_{\sigma_z}(t)] dt$ as close to zero as possible.
Let us examine a special situation where we can actually reduce the term $\int_{0}^{t}\Delta(t)[M_{\sigma_z}(t)] dt$ completely to zero. We can take advantage of the fact that as long as $g$ is 
a constant, 
$M_{\sigma_z}(t)$ is periodic with a period $\frac{2\pi}{\omega}$. Keeping $g$ a constant, first we keep $\Delta$ at a certain nonzero value for a $k\frac{2\pi}{\omega}$ period of 
time ($k$: integer), 
then we flip $\Delta$ to the opposite sign for another $k\frac{2\pi}{\omega}$ period of time. As a result, the two parts $\int_{0}^{k\frac{2\pi}{\omega}}\Delta(t)[M_{\sigma_z}(t)] dt$ and $\int_{k\frac{2\pi}{\omega}}^{2k\frac{2\pi}{\omega}}\Delta(t)[M_{\sigma_z}(t)] dt$ cancel each other, eliminating the first-order dephasing effect of $\Delta$ for the duration of $(0,2k\frac{2\pi}{\omega})$.

\begin{figure}
\includegraphics[width=0.5\textwidth]{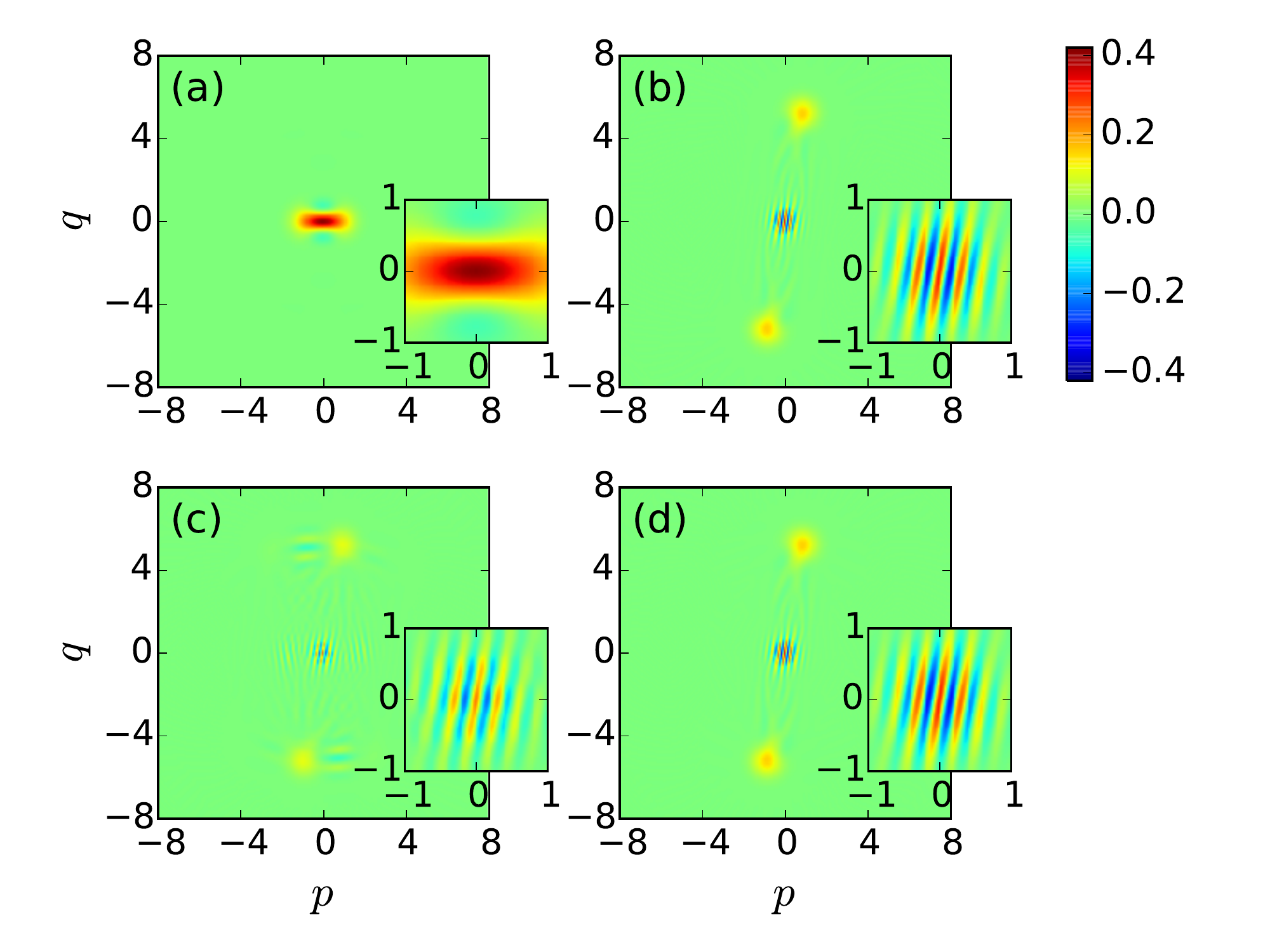}
\caption{\label{fig:CatGrowthAndPreservation}
(color online)
Cat-state amplification and rephasing in the Wigner representation of the oscillator state projected onto the qubit state 
$\ket{+}_x+\ket{-}_x$.
The $x$ and $y$ axes are the oscillator's dimensionless field quadratures $p$ and $q$. 
The parameters are modulated as explained in the text with $\Delta=0.1\omega$ and $g(0)=0.833\omega$. The insets show the central parts of the figures. 
(a) The initial state, which is taken to be the ground state, reasonably resembles the cat state 
$\ket{-\frac{g(0)}{\omega}} + \ket{+\frac{g(0)}{\omega}}$, 
with a fidelity of 
0.99986. 
After the initial state goes through two oscillator periods of the sinusoidally driven cat-state-amplification process, the resulting state is shown in (b), which resembles the target cat state 
$\ket{-\frac{\tilde{g}(4\pi/\omega)}{\omega}} + \ket{+\frac{\tilde{g}(4\pi/\omega)}{\omega}}$ 
with $|\tilde{g}/\omega|=5.3$ and a fidelity of 
0.989. 
We now let the state in (b) freely evolve for ten oscillator periods, and the resulting state of the oscillator is shown in (c).
The fidelity between 
the state of (c) and the target cat state is only
0.933, 
since the state is dephased by the nonzero $\Delta$.
However, if we insert one $\hat{\sigma}_x$-$\pi$ pulse in the middle 
of the ten oscillator periods, the evolved state, shown in (d), has a 
0.989 
fidelity with the target cat state, 
analogous to a Hahn spin-echo rephasing effect.
	}
\end{figure}

\section{$\pi$-pulses and their applications}
Above we have discussed 
the method of eliminating the first-order effect of finite $\Delta$ by flipping the sign of $\Delta$.
Here we will show a scheme to manipulate $g$ or $\Delta$ indirectly.
By applying $\pi$-pulses to the qubit alone, which is a commonly used technique in dynamical decoupling \cite{viola1999viola, de2010g}, we can achieve the effect of flipping the signs of $g$ or $\Delta$.
There are three basic types of $\pi$-pulses, each of which amounts to applying a Pauli operator ($\hat{\sigma}_z$, $\hat{\sigma}_x$ or $\hat{\sigma}_y$) to the qubit.
Let us first examine the $\hat{\sigma}_z$-$\pi$-pulse. Since $\hat{\sigma}_x$ and $\hat{\sigma}_z$ anticommute, we have
$\hat{H}(g)\hat{\sigma}_z = \hat{\sigma}_z\hat{H}(-g)$ in Eq.~(\ref{eqn:Hamiltonian_full}).
Therefore, without directly altering the qubit-oscillator coupling coefficient $g$, in the Hamiltonian the sign of $g$ is flipped by applying $\hat{\sigma}_z$-$\pi$ pulses to the qubit alone. 
For $\hat{\sigma}_x$ and $\hat{\sigma}_y$ pulses, similar arguments apply.
By applying $\hat{\sigma}_y$-$\pi$ pulses to the qubit, the signs of both $g$ and $\Delta$ are simultaneously flipped. 
As explained below, cat states can be amplified by applying $\sigma_z$-$\pi$ pulses on the qubit, and dephasing due to $\Delta$ can be corrected by $\sigma_x$-$\pi$ pulses. Both amplifying a cat state and rephasing can be achieved by $\sigma_y$-$\pi$ pulses. See Appendix D for details.
\section{Numerical simulations}
When $\Delta$ is nonzero but still small compared to $\omega$, the ground state, which can be written as $\ket{E_{0+}(0)}=\ket{+}_x\ket{\psi_-(0)}+\ket{-}_x\ket{\psi_+(0)}$, is very close to $\ket{E^{(0)}_{0+}(0)}=\ket{+}_x\ket{-\frac{g(0)}{\omega}}+\ket{-}_x\ket{+\frac{g(0)}{\omega}}$ [Eq.~(\ref{eqn:energy_eigenstates})]. 
Although $\ket{\psi_-(0)}+\ket{\psi_+(0)}$ is not strictly a cat state of two coherent states, 
the fidelity of $\ket{\psi_-(0)}+\ket{\psi_+(0)}$ to the cat state $\ket{-\frac{g(0)}{\omega}}+\ket{+\frac{g(0)}{\omega}}$ is very high as shown in the caption of Fig.~\ref{fig:CatGrowthAndPreservation}(a).

Amplifying a cat state can be achieved either by directly modulating $g(t)$, or by indirectly flipping the sign of $g(t)$ by applying $\hat{\sigma}_z$-$\pi$ pulses to the qubit alone, as shown above. Here, we show an example of the former scheme. See Appendix D for the latter scheme.

We begin with the ground state $\ket{E_{0+}(0)}$ with constant $g(0)$ as the initial state. 
The Wigner representation of its oscillator part $\ket{\psi_-(0)}+\ket{\psi_+(0)}$ is shown in Fig.~\ref{fig:CatGrowthAndPreservation}(a).
We apply a short sinusoidal driving force by modulating
$g(t)=g(0)\cos\omega t$
for two periods of the oscillator ($t=0$ to $t=\frac{4\pi}{\omega}$) as shown in Figs.~\ref{fig:gAndgTilde}(c) and \ref{fig:gAndgTilde}(d). 
During this modulation, $\tilde{g}$ evolves as shown in Figs.~\ref{fig:gAndgTilde}(c) and \ref{fig:gAndgTilde}(d). 
By the end of the modulation ($t=\frac{4\pi}{\omega}$), the amplitude of the cat state is amplified by 
a factor of 6.4.
The Wigner representation of the resulting state $\ket{\psi_-(4\pi/\omega)}+\ket{\psi_+(4\pi/\omega)}$, which has a high fidelity with 
$\ket{-\frac{\tilde{g}(4\pi/\omega)}{\omega}} + \ket{+\frac{\tilde{g}(4\pi/\omega)}{\omega}}$, 
is shown in Fig.~\ref{fig:CatGrowthAndPreservation}(b).
We can repeat this process to further increase the absolute value of amplitudes of the oscillator states, if we want. 
Note that the amplified cat in Fig.~\ref{fig:CatGrowthAndPreservation}(b) is tilted by an angle of about $\pi/2$ relative to the $x$-axis. This is due to the fact that $\tilde{g}(4\pi/\omega)$ lags $g(4\pi/\omega)$ by about a quarter of a cycle, as indicated by the dots in Fig.~\ref{fig:gAndgTilde}(c) and \ref{fig:gAndgTilde}(d).

The size of the cat state in Fig.~\ref{fig:CatGrowthAndPreservation}(b), quantified by the distance between the two coherent states in the superposition, is comparable to that of Ref.~\cite{vlastakis2013deterministically}, i.e., $\sim 100$ photons. In Appendix C, we show simulations for an effectively 400-photon cat state. Our cat-amplification scheme also works for weaker coupling, as we show with an example in Appendix D.

Now we
let the state shown in Fig.~\ref{fig:CatGrowthAndPreservation}(b) evolve freely for $10(2\pi/\omega)$, then the state is eventually dephased, as shown in Fig.~\ref{fig:CatGrowthAndPreservation}(c). 
To counter this, we instead in Fig.~\ref{fig:CatGrowthAndPreservation}(d) allow the cat state to evolve freely for five oscillator periods, apply one $\sigma_x$-$\pi$ pulse to the qubit, and then allow free evolution for another five periods.
In Fig.~\ref{fig:CatGrowthAndPreservation}(d), we can see that the amplified cat state is recovered, in analogy to the rephasing of the Hahn spin-echo method~\cite{Hahn50PR}.
To preserve a state for a longer time, more $\hat{\sigma}_x$-$\pi$ pulses can be applied in the same manner. Also we can apply $\hat{\sigma}_x$-$\pi$ pulses more frequently to counter the dephasing by a larger $\Delta$.

Throughout this paper, we have neglected the effects of dissipations, since our protocol can be completed in the time scale of the oscillator's period ($\sim1$ ns), which is much shorter than the decoherence time ($\sim100$ ns) determined by the qubit and the $\sim100$-photon state. Here we assume the single photon decay time of $\sim10$ $\rm{\mu}$s, the qubit decoherence time of $\sim1$ $\rm{\mu}$s, and a 10 GHz oscillator with parameters used in Fig.~\ref{fig:CatGrowthAndPreservation}.

\section{Conclusion}
We have studied the evolution of the quantum state in a qubit-oscillator system with a time-dependent coupling, in the case of a small qubit frequency compared to the oscillator frequency. 
We have analytically shown that the quantum state evolution can be simply expressed by introducing the dynamical evolution eigenstates.
Using this method, we have designed a scheme for generating large cat states that is orders of magnitude faster than known methods \cite{vlastakis2013deterministically}, 
and a scheme to rephase the cat states in the oscillator using $\pi$ pulses on the qubit, which is entangled with the oscillator.
We point out that our method is quite general and can be used for general entangled-cat-state engineering with a wide range of the system parameters in various systems. 
Also, our techniques of $\pi$ pulses can be widely used to protect quantum information in recently advancing cavity- and circuit-QED systems.

A part of simulations in this study was performed using QuTiP \cite{qutip}. 

\begin{acknowledgments}
T.F., F.Y., K.S., M.S., and M.T. would like to
acknowledge support from Japan Science and Technology Agency Core Research for Evolutionary Science and Technology (Grant No. JPMJCR1775).
Z.X and J.P.D. would like to acknowledge AFOSR, ARO, DARPA, NSF, and NGAS.
\end{acknowledgments}

\appendix

\section{Derivation of the state evolution with infinitesimal qubit frequency}
Under the  limit of $\Delta/\omega\rightarrow0$, 
the Hamiltonian of the combined system of the qubit and the oscillator is

\begin{equation} \begin{split}
\hat{H}^{(0)}(t)=\hbar\omega\left(\hat{a}^\dagger\hat{a}+\frac{1}{2}\right)+\hbar g(t)\hat{\sigma}_x(\hat{a}^\dagger+\hat{a}).
\end{split}\end{equation}


Let us examine the evolution of an initial state in the form of:
\begin{equation} \begin{split}
\ket{\Phi_{N\pm}(0)}=\ket{\pm}_x\hat{D}\left({\mp\frac{\tilde{g}(0)}{\omega}}\right)\ket{N}.
\end{split}\end{equation} 
We can choose particular $\pm$ and $N$, but the following calculation applies to all $\pm$ and $N$. We can also see that with various choices of $N$ and $\pm$, $\ket{\Phi_{N\pm}(0)}$ represents a complete orthonormal basis regardless of the value of $\tilde{g}(0)$. By examining the evolution of initial state $\ket{\Phi_{N\pm}(0)}$ with all choices of $\pm$ and $N$, we can understand the evolution of any general initial state, which itself can be expressed as a superposition of $\ket{\Phi_{N\pm}(0)}$ with different $\pm$ and $N$. Also note that we can set $\tilde{g}(0)$ to any value.

To calculate the evolution of the initial state $\ket{\Phi_{N\pm}(0)}$ to the final time $t_\text{F}$, we divide the time period of interest into $K+1$ small segments, so that the entire period is divided by time points $0, t_1, t_2, \cdots , t_i, t_{i+1}, \cdots, t_K, t_\text{F}$. Each segment is considered to be small enough so that, in a single segment $g(t)$ and $\hat{H}^{(0)}(t)$ do not change much and are treated as constants. Therefore the final state at $t=t_\text{F}$ can be expressed as
\begin{equation}\begin{split}
\ket{\Phi_{N\pm}(t_\text{F})}=&\exp [-i\hat{H}^{(0)}(t_K)\times(t_\text{F}-t_K)/\hbar] \cdots\\
&\times \exp [-i\hat{H}^{(0)}(t_j)\times(t_{j+1}-t_j)/\hbar] \cdots\\
&\times \exp [-i\hat{H}^{(0)}(0)\times t_1/\hbar]\\ 
&\times \ket{\pm}_x\hat{D}\left({\mp\frac{\tilde{g}(0)}{\omega}}\right)\ket{N}.
\end{split}\end{equation} 

Now let us solve for $\ket{\Phi_{N\pm}(t_\text{F})}$. At each time point $t=t_j$, starting with $t=0$,
we carry out the following procedures:

(1) We make the ansatz that, at any time point $t=t_j$, the evolved state is in the form of
\begin{equation}\begin{split}
\ket{\Phi_{N\pm}(t_j)}=e^{i\phi_{N}(t_j)}\ket{\pm}_x\hat{D}\left({\mp\frac{\tilde{g}(t_j)}{\omega}}\right)\ket{N},
\end{split}\end{equation} 
where $\ket{\pm}$ and $\ket{N}$ are the same as the initial state $\ket{\Phi_{N\pm}(0)}$. This is obviously true at $t=0$ and we will show, by mathematical induction, that indeed at each time point the state can be expressed in such a form. The complex number $\tilde{g}(t_j)$
generally changes at different time points, and a phase $e^{i\phi_{N\pm}(t_j)}$ can accumulate as well. 

(2) 
To make the expression more compact, let $\tau_j=t_{j+1}-t_j$, $g_j=g(t_j)$, $\tilde{g}_j=\tilde{g}(t_j)$. 
The state at $t_{j+1}$, evolving from the state at the previous time point $t_{j}$, can be expressed as follows:
\begin{widetext}
\begin{equation}\label{eqn:StepOfState1}\begin{split}
\ket{\Phi_{N\pm}(t_{j+1})}&=\exp [-i\hat{H}^{(0)}(t_j)(t_{j+1}-t_j)/\hbar]\ket{\Phi_{N\pm}(t_{j})}\\
&=\exp \left\{-i\left[\omega\left(\hat{a}^\dagger\hat{a}+\frac{1}{2}\right)\pm g_j(\hat{a}^\dagger+\hat{a})\right]\tau_j\right\} \exp [i\phi_{N}(t_j)]\hat{D}\left({\mp\frac{\tilde{g}_j}{\omega}}\right)\ket{\pm}_x\ket{N}\\
&=\exp [i\phi_{N}(t_j)]\exp \{-i\Im[(\tilde{g}_j-g_j)g_j/\omega^2]\}\exp \left\{-i\left[\omega\left(\hat{a}^\dagger\hat{a}+\frac{1}{2}\right)\pm g_j(\hat{a}^\dagger+\hat{a})\right]\tau_j\right\}\\
&\quad \times \hat{D}\left({\mp\frac{\tilde{g}_j-g_j}{\omega}}\right)\hat{D}\left({\mp\frac{g_j}{\omega}}\right)\ket{\pm}_x\ket{N}\\
&=\exp [i\phi_{N}(t_j)]\exp \{-i\Im[(\tilde{g}_j-g_j)g_j/\omega^2]\}\exp \left\{-i\left[\omega\left(\hat{a}^\dagger\hat{a}+\frac{1}{2}\right)\pm g_j(\hat{a}^\dagger+\hat{a})\right]\tau_j\right\}\\
&\quad \times\exp \left\{\left(\mp\frac{\tilde{g}_j-g_j}{\omega}\right)\left[\left(\hat{a}^\dagger\pm \frac{g_j}{\omega}\right)\mp \frac{g_j}{\omega}\right]-\left(\mp\frac{\tilde{g}^*_j-g_j}{\omega}\right)\left[\left(\hat{a}\pm \frac{g_j}{\omega}\right)\mp \frac{g_j}{\omega}\right]\right\}\hat{D}\left({\mp\frac{g_j}{\omega}}\right)\ket{\pm}_x\ket{N}\\
&=\exp [i\phi_{N}(t_j)]\exp \{-i\Im[(\tilde{g}_j-g_j)g_j/\omega^2]\}\\
&\quad \times\exp \left\{\left(\mp\frac{\tilde{g}_j-g_j}{\omega}\right)\left[\left(\hat{a}^\dagger\pm \frac{g_j}{\omega}\right)e^{-i\tau_j\omega}\mp \frac{g_j}{\omega}\right]-\left(\mp\frac{\tilde{g}^*_j-g_j}{\omega}\right)\left[\left(\hat{a}\pm \frac{g_j}{\omega}\right)e^{+i\tau_j\omega}\mp \frac{g_j}{\omega}\right]\right\}\\
&\quad\times \exp \left\{-i\left[\omega\left(\hat{a}^\dagger\hat{a}+\frac{1}{2}\right)\pm g_j(\hat{a}^\dagger+\hat{a})\right]\tau_j\right\}\hat{D}\left({\mp\frac{g_j}{\omega}}\right)\ket{\pm}_x\ket{N}\\
&=\exp [i\phi_{N}(t_j)]\exp \{-i\Im[(\tilde{g}_j-g_j)g_j/\omega^2]\}\hat{D}\left(\mp\frac{\tilde{g}_j-g_j}{\omega}e^{-i\tau_j\omega}\right)\\
&\quad \times\exp\left[-\frac{(\tilde{g}_j-g_j)g_j}{\omega^2}e^{-i\tau_j\omega}+\frac{(\tilde{g}^*_j-g_j)g_j}{\omega^2}e^{+i\tau_j\omega}+\frac{(\tilde{g}_j-g_j)g_j}{\omega^2}-\frac{(\tilde{g}^*_j-g_j)g_j}{\omega^2}\right]\\
&\quad \times\exp \left\{-i\left[\omega{\left(N+\frac{1}{2}\right)}-\frac{g^2_i}{\omega}\right]\tau_j\right\}\hat{D}\left({\mp\frac{g_j}{\omega}}\right)\ket{\pm}_x\ket{N}\\
&=\exp [i\phi_{N}(t_j)]\exp \left\{i\Im\left[\frac{(\tilde{g}_j-g_j)g_j}{\omega^2}(1-e^{-i\tau_j\omega})\right]\right\}\\
&\quad \times \exp \left\{-i\left[\omega{\left(N+\frac{1}{2}\right)}-\frac{g^2_i}{\omega}\right]\tau_j\right\}\hat{D}\left({\mp\frac{g_j}{\omega}\mp\frac{\tilde{g}_j-g_j}{\omega}e^{-i\tau_j\omega}}\right)\ket{\pm}_x\ket{N},\\
\end{split}\end{equation}
where the symbol $\Im$ means an imaginary part. In the second to the last step of Eq.~(\ref{eqn:StepOfState1}), we have used
\begin{equation}\label{eqn:}\begin{split}
&\left[\omega\left(\hat{a}^\dagger\hat{a}+\frac{1}{2}\right)\pm g_j(\hat{a}^\dagger+\hat{a})\right]\hat{D}({\mp\frac{g_j}{\omega}})\ket{\pm}_x\ket{N}\\
&\quad =\hat{D}\left({\mp\frac{g_j}{\omega}}\right)\bigg[\omega\left(\hat{a}^\dagger\mp\frac{g_j}{\omega}\right)\left(\hat{a}\mp\frac{g_j}{\omega}\right)
\pm g_j\left(\hat{a}^\dagger\mp\frac{g_j}{\omega}+\hat{a}\mp\frac{g_j}{\omega}\right)+\frac{1}{2}\omega\bigg]\ket{\pm}_x\ket{N}\\
&\quad =\hat{D}\left({\mp\frac{g_j}{\omega}}\right)\left[\omega\left(\hat{a}^\dagger\hat{a}+\frac{1}{2}\right)-\frac{g^2_i}{\omega}\right]\ket{\pm}_x\ket{N}
 =\hat{D}\left({\mp\frac{g_j}{\omega}}\right)\left[\omega{\left(N+\frac{1}{2}\right)}-\frac{g^2_i}{\omega}\right]\ket{\pm}_x\ket{N}\\
&\quad =\left[\omega{\left(N+\frac{1}{2}\right)}-\frac{g^2_i}{\omega}\right]\hat{D}\left({\mp\frac{g_j}{\omega}}\right)\ket{\pm}_x\ket{N},
\end{split}\end{equation}
and in the third to the last step we have used
\begin{equation}\label{eqn:}\begin{split}
\exp\left\{-i\tau_j\left[\omega\left(\hat{a}^\dagger\hat{a}+\frac{1}{2}\right)\pm g_j\left(\hat{a}^\dagger+\hat{a}\right)\right]\right\}\left(\hat{a}\pm \frac{g_j}{\omega}\right)
=\left(\hat{a}\pm \frac{g_j}{\omega}\right)e^{i\tau_j\omega}\exp\left\{-i\tau_j\left[\omega\left(\hat{a}^\dagger\hat{a}+\frac{1}{2}\right)\pm g_j\left(\hat{a}^\dagger+\hat{a}\right)\right]\right\}.
\end{split}\end{equation}
Now recall that
\begin{equation}\begin{split}
\ket{\Phi_{N\pm}(t_j)}=\exp [i\phi_{N}(t_j)] \ket{\pm}_x\hat{D}\left({\mp\frac{\tilde{g}(t_j)}{\omega}}\right)\ket{N}.
\end{split}\end{equation} 
Therefore we have proved, by mathematical induction, that the quantum state at $t=t_{j+1}$ also has the form
\begin{equation}\begin{split}
\ket{\Phi_{N\pm}(t_{j+1})}=\exp [i\phi_{N}(t_{j+1})] \ket{\pm}_x\hat{D}\left({\mp\frac{\tilde{g}(t_{j+1})}{\omega}}\right)\ket{N},
\end{split}\end{equation}
with
\begin{equation}\begin{split}
\tilde{g}(t_{j+1})=g(t_{j})+[\tilde{g}(t_{j})-g(t_{j})]\exp [-i\omega(t_{j+1}-t_{j})]
\end{split}\end{equation}
and 
\begin{equation}\begin{split}
\phi_{N}(t_{j+1})=\phi_{N}(t_j)+\frac{\Im\{(\tilde{g}^*_j-g_j) g_j[1-{\exp (-i\omega\tau_j)}]\}}{\omega^2}-\left(N+\frac{1}{2}\right)\omega\tau_j+\frac{g^2_i}{\omega}\tau_j.
\end{split}\end{equation}

Since $t_{j+1}-t_{j}=\tau_j\rightarrow0$, we have
\begin{equation}\begin{split}
\tilde{g}(t_{j+1})=\tilde{g}(t_{j})[1-i\omega(t_{j+1}-t_{j})]-g(t_{j})[-i\omega(t_{j+1}-t_{j})]
\end{split}\end{equation}
and
\begin{equation}\begin{split}
\phi_{N}\left(t_{j+1}\right)=\phi_{N}(t_j) +\frac{\Im[i(\tilde{g}^*_j-g_j) g_j ]}{\omega} \tau_j
-(N+\frac{1}{2})\omega\tau_j+\frac{g^2_i}{\omega}\tau_j.
\end{split}\end{equation}
Therefore,
\begin{equation}\begin{split}
\frac{\tilde{g}(t_{j+1})-\tilde{g}(t_{j})}{t_{j+1}-t_{j}}=i\omega[g(t_{j})-\tilde{g}(t_{j})],
\end{split}\end{equation}
which leads to
\begin{equation}\label{eqn:differential_g} \begin{split}
\frac{d}{d t}\tilde{g}(t)=i \omega (g(t)-\tilde{g}(t)),
\end{split}\end{equation}
for which the solution is
\begin{equation}\label{eqn:gTildeT} \begin{split}
\tilde{g}(t)=\tilde{g}(0) e^{-i \omega t}+ e^{-i \omega t} \int_{0}^{t}i \omega g(t')e^{i \omega t'}\rm{d} t'.
\end{split}\end{equation}
As the state evolves under changing $g$, the phase $\phi_{N}(t_\text{F})$ will accumulate and the final state is 
\begin{equation}\label{eqn:phiT}\begin{split}
\ket{\Phi_{N\pm}(t_\text{F})}=e^{i\phi_{N}(t_\text{F})}\hat{D}\left({\mp\frac{\tilde{g}(t_\text{F})}{\omega}}\right)\ket{\pm}_x\ket{N}.
\end{split}\end{equation} 
where $\tilde{g}=\tilde{g}(t)$ is given by Eq.~(\ref{eqn:gTildeT}) and $\phi_{N}(t_\text{F})=\int_{0}^{t_\text{F}}\left(\frac{\Im[i(\tilde{g}^*-g) g]}{\omega}+\frac{g^2}{\omega}-\left(N+\frac{1}{2}\right)\omega\right)dt$.
%
Notice in the phase $\phi_{N}(t_\text{F})$, the part of $\int_{0}^{t_\text{F}}\left(\frac{\Im[i(\tilde{g}^*-g) g]}{\omega}+\frac{g^2}{\omega}-\frac{\omega}{2}\right)dt$ is the same for every $N$ and $\pm$, therefore we can simplify it as $\phi_{N}(t_\text{F})=-N\omega t_\text{F}$. Combining $\ket{\Phi_{N+}(t_\text{F})}$ and $\ket{\Phi_{N-}(t_\text{F})}$, we arrive at Eqs.~(4) and (5) in the main text.
\newline
\section{Derivation of the state evolution with nonzero qubit frequency}

We now consider the case where $\Delta$ is not infinitesimal and use the full Hamiltonian  $\hat{H}(t)=\hat{H}^{(0)}(t)-\frac{\hbar}{2}\Delta\hat{\sigma}_z$. As explained above, the dynamical energy eigenstates form a complete basis.
Therefore, at any time $t$, we can express any quantum state of interest as a superposition of dynamical energy eigenstates: $\ket{\varphi(t)}=\sum_{N,\pm}^{}C_{N\pm}(t)\ket{\tilde{E}^{(0)}_{N\pm}(t)}$.
As the state $\ket{\varphi(t)}$ evolves, so does every dynamical energy eigenstate $\ket{\tilde{E}^{(0)}_{N\pm}(t)}$ and its corresponding coefficient $C_{N\pm}(t)$.
Now we need to solve the evolution of the coefficients $C_{N\pm}(t)$ to completely determine the evolution of state $\ket{\varphi(t)}$. Starting from initial time zero, we divide the time period of interest $(0,t)$ into $K+1$ segments: ($0, t_1, t_2, ... , t_K, t$),
with each segment being infinitesimal. For convenience we will adopt the matrix form, therefore the state at time $t$, which evolves from the initial state $\ket{\varphi(0)}$ under Hamiltonian $\hat{H}(t)$, can be written as:

\begin{equation} \begin{split}\label{eqn:final_state_SmallDelta_AllOrder_MatrixForm}
\ket{\varphi(t)}=&\sum_{N,\pm}^{}C_{N\pm}(t)e^{-i N\omega t}\ket{\tilde{E}^{(0)}_{N\pm}(t)}\\
=&\begin{bmatrix}
C_{0+}(t)       & C_{1+}(t)  &   \dots & C_{0-}(t)  & C_{1-}(t)  &  \dots 
\end{bmatrix}
\begin{bmatrix}
\ket{\tilde{E}^{(0)}_{0+}(t)} & e^{-i \omega t}\ket{\tilde{E}^{(0)}_{1+}(t)} &  \dots & \ket{\tilde{E}^{(0)}_{0-}(t)} & e^{-i \omega t}\ket{\tilde{E}^{(0)}_{1-}(t)} \dots 
\end{bmatrix}^T\\
=&\exp[-i\hat{H}^{}(t_K)\times(t-t_K)/\hbar]  \cdots  \exp[-i\hat{H}^{}(t_1)\times(t_2-t_1)/\hbar]\exp[-i\hat{H}^{}(0)\times t_1/\hbar]\\
&\times\begin{bmatrix}
C_{0+}(0)       & C_{1+}(0)  &   \dots & C_{0-}(0)  & C_{1-}(0)  &  \dots 
\end{bmatrix}\cdot
\begin{bmatrix}
\ket{\tilde{E}^{(0)}_{0+}(0)} & \ket{\tilde{E}^{(0)}_{1+}(0)} &  \dots & \ket{\tilde{E}^{(0)}_{0-}(0)} &\ket{\tilde{E}^{(0)}_{1-}(0)} \dots 
\end{bmatrix}^T.
\end{split}\end{equation} 
From $t_{j}$ to $t_{j+1}$, so long as $t_{j+1}-t_{j}\rightarrow 0$, the state evolves in the following way:
\begin{equation} \begin{split}\label{eqn:final_state_SmallDelta_Induction}
\ket{\varphi(t_{j+1})}=&\exp\left\{-i\left[\hat{H}^{(0)}(t_{j})-\frac{\hbar}{2}\Delta\hat{\sigma}_z\right](t_{j+1}-t_{j})/\hbar\right\}\ket{\varphi(t_{j})}\\
=&\exp\left\{-i\left[\hat{H}^{(0)}(t_{j})-\frac{\hbar}{2}\Delta\hat{\sigma}_z\right](t_{j+1}-t_{j})/\hbar\right\}\sum_{N,\pm}^{}C_{N\pm}(t_{j})e^{-i N\omega t_{j}}\ket{\tilde{E}^{(0)}_{N\pm}(t_{j})}\\
=&\sum_{N,\pm}^{}C_{N\pm}(t_{j})\exp\left[-i\left(-\frac{1}{2}\Delta\hat{\sigma}_z\right)(t_{j+1}-t_{j})\right]\exp[-i\hat{H}^{(0)}(t_{j})\times(t_{j+1}-t_{j})/\hbar]e^{-i N\omega t_{j}}\ket{\tilde{E}^{(0)}_{N\pm}(t_{j})}\\
=&\sum_{N,\pm}^{}C_{N\pm}(t_{j})\exp\left[-i\left(-\frac{1}{2}\Delta\hat{\sigma}_z\right)(t_{j+1}-t_{j})\right]e^{-i N\omega (t_{j+1}-t_{j})}e^{-i N\omega t_{j}}\ket{\tilde{E}^{(0)}_{N\pm}(t_{j+1})}\\
=&\sum_{N,\pm}^{}\sum_{N',\pm}^{}C_{N\pm}(t_{j})e^{-i N'\omega t_{j+1}}\ket{\tilde{E}^{(0)}_{N'\pm}(t_{j+1})}\\
&\times \bra{\tilde{E}^{(0)}_{N'\pm}(t_{j+1})}e^{+i N'\omega t_{j+1}}\exp\left[-i\left(-\frac{1}{2}\Delta\hat{\sigma}_z\right)(t_{j+1}-t_{j})\right]e^{-i N\omega t_{j+1}}\ket{\tilde{E}^{(0)}_{N\pm}(t_{j+1})}\\
=&\begin{bmatrix}
C_{0+}(t_j)       & C_{1+}(t_j)  &   \dots & C_{0-}(t_j)  & C_{1-}(t_j)  &  \dots 
\end{bmatrix}
\exp[-i(-\frac{1}{2}\Delta) [M_{\sigma_z}(t_{j+1})](t_{j+1}-t_{j})]\\
&\times\begin{bmatrix}
\ket{\tilde{E}^{(0)}_{0+}(t_{j+1})} & e^{-i \omega t_{j+1}}\ket{\tilde{E}^{(0)}_{1+}(t_{j+1})} &  \dots & \ket{\tilde{E}^{(0)}_{0-}(t_{j+1})} & e^{-i \omega t_{j+1}}\ket{\tilde{E}^{(0)}_{1-}(t_{j+1})} \dots 
\end{bmatrix}^T,
\end{split}\end{equation} 
where we have used the relation 
\begin{equation} \begin{split}\label{}
&\exp\left \{-i\left[\hat{H}^{(0)}(t_{j})-\frac{\hbar}{2}\Delta\hat{\sigma}_z\right](t_{j+1}-t_{j})/\hbar\right \}\\
&\quad =\exp\left[-i\left (-\frac{\hbar}{2}\Delta\hat{\sigma}_z\right )(t_{j+1}-t_{j})/\hbar\right]
\exp[-i\hat{H}^{(0)}(t_{j})\times(t_{j+1}-t_{j})/\hbar]\exp[\mathcal{O}(t_{j+1}-t_{j})^2]\\
&\quad \xrightarrow[]{t_{j+1}-t_{j}\rightarrow 0}\exp\left[-i\left (-\frac{\hbar}{2}\Delta\hat{\sigma}_z\right )(t_{j+1}-t_{j})/\hbar\right]\exp[-i\hat{H}^{(0)}(t_{j})\times(t_{j+1}-t_{j})/\hbar].
\end{split}\end{equation} 
The term $M_{\sigma_z}(t)$ in Eq.~(\ref{eqn:final_state_SmallDelta_Induction}) is the matrix expansion of operator $\hat{\sigma}_z$ in the basis $\ket{\tilde{E}^{(0)}_{N\pm}(t)}$ at time $t$: 

	\begin{equation} \begin{split}\label{eqn:final_state_SmallDelta_M_Matrix}
	M_{\sigma_z}(t)=
	\begin{bmatrix}
	M_{\sigma_z}^{(+,+)}(t)       & M_{\sigma_z}^{(-,+)}(t)\\
	M_{\sigma_z}^{(+,-)}(t)       & M_{\sigma_z}^{(-,-)}(t)\\
	\end{bmatrix},
	\end{split}\end{equation} 
where
\begin{equation} \begin{split}\label{eqn:final_state_SmallDelta_M_Matrix_block_pmpm}
M_{\sigma_z}^{(\pm,\pm)}(t)=\begin{bmatrix}
\bra{\tilde{E}^{(0)}_{0\pm}(t)}\hat{\sigma}_z\ket{\tilde{E}^{(0)}_{0\pm}(t)}       & \bra{\tilde{E}^{(0)}_{1\pm}(t)}\hat{\sigma}_z\ket{\tilde{E}^{(0)}_{0\pm}(t)}e^{+i \omega t} &   \dots\\
\bra{\tilde{E}^{(0)}_{0\pm}(t)}\hat{\sigma}_z\ket{\tilde{E}^{(0)}_{1\pm}(t)}e^{-i \omega t}       & \bra{\tilde{E}^{(0)}_{1\pm}(t)}\hat{\sigma}_z\ket{\tilde{E}^{(0)}_{1\pm}(t)} &   \dots\\
\hdotsfor{3} \\
\end{bmatrix},
\end{split}\end{equation} 
and
	\begin{equation} \begin{split}\label{eqn:final_state_SmallDelta_M_Matrix_block_pmmp}
	M_{\sigma_z}^{(\pm,\mp)}(t)=\begin{bmatrix}
	\bra{\tilde{E}^{(0)}_{0\pm}(t)}\hat{\sigma}_z\ket{\tilde{E}^{(0)}_{0\mp}(t)}       & \bra{\tilde{E}^{(0)}_{1\pm}(t)}\hat{\sigma}_z\ket{\tilde{E}^{(0)}_{0\mp}(t)}e^{+i \omega t} &   \dots\\
	\bra{\tilde{E}^{(0)}_{0\pm}(t)}\hat{\sigma}_z\ket{\tilde{E}^{(0)}_{1\mp}(t)}e^{-i \omega t}       & \bra{\tilde{E}^{(0)}_{1\pm}(t)}\hat{\sigma}_z\ket{\tilde{E}^{(0)}_{1\mp}(t)} &   \dots\\
	\hdotsfor{3} \\
	\end{bmatrix}.
	\end{split}\end{equation} 

On the other hand,
\begin{equation} \begin{split}\label{eqn:final_state_SmallDelta_Induction1}
\ket{\varphi(t_{j+1})}=&\sum_{N,\pm}^{}C_{N\pm}(t_{j+1})e^{-i N\omega t_{j+1}}\ket{\tilde{E}^{(0)}_{N\pm}(t_{j+1})}\\
=&\begin{bmatrix}
C_{0+}(t_{j+1})       & C_{1+}(t_{j+1})  &   \dots & C_{0-}(t_{j+1})  & C_{1-}(t_{j+1})  &  \dots 
\end{bmatrix}\\
&\times\begin{bmatrix}
\ket{\tilde{E}^{(0)}_{0+}(t_{j+1})} & e^{-i \omega t_{j+1}}\ket{\tilde{E}^{(0)}_{1+}(t_{j+1})} &  \dots & \ket{\tilde{E}^{(0)}_{0-}(t_{j+1})} & e^{-i \omega t_{j+1}}\ket{\tilde{E}^{(0)}_{1-}(t_{j+1})} \dots 
\end{bmatrix}^T.
\end{split}\end{equation} 

Comparing Eqs.~(\ref{eqn:final_state_SmallDelta_Induction}) and  (\ref{eqn:final_state_SmallDelta_Induction1}) and using mathematical induction, we have
\begin{equation} \begin{split}\label{eqn:final_state_SmallDelta_coefficients_row}
&\begin{bmatrix}
C_{0+}(t)       & C_{1+}(t)  &   \dots & C_{0-}(t)  & C_{1-}(t)  &  \dots 
\end{bmatrix}
=\begin{bmatrix}
C_{0+}(0)       & C_{1+}(0)  &   \dots & C_{0-}(0)  & C_{1-}(0)  &  \dots 
\end{bmatrix} J(0,t),
\end{split}\end{equation} 
where

\begin{equation}\begin{split}\label{eqn:final_state_SmallDelta_coefficients_evolution_matrix}
J(0,t)=\exp\left[-i\left(-\frac{1}{2}\Delta\right)[M_{\sigma_z}(t_1)]t_1\right]
\exp\left[-i\left(-\frac{1}{2}\Delta\right) [M_{\sigma_z}(t_2)](t_2-t_1)\right]\cdots
\exp\left[-i\left(-\frac{1}{2}\Delta\right) [M_{\sigma_z}(t)](t-t_K)\right].
\end{split}\end{equation}

Note that, generally, at different time points $t$ and $t'$, $M_{\sigma_z}(t)$ and $M_{\sigma_z}(t')$ do not commute, making the analytical calculation of the evolution matrix $J(0,t)$ very complicated. But in the case when we only consider up to the first-order effect of small $\Delta$, we can ignore the second order $\Delta$ term, which means we consider the commutation 
$[\Delta\times M_{\sigma_z}(t), \Delta\times M_{\sigma_z}(t')]\sim O(\Delta^2)\sim 0$, 
and therefore 
$\Delta\times M_{\sigma_z}(t)$ and $\Delta\times M_{\sigma_z}(t')$ 
approximately commute. This enables us to calculate the evolution matrix $J(0,t)$ and state $\ket{\varphi(t)}$ up to the first order:
\begin{equation} \begin{split}
J^{(1)}(0,t)=\exp\left[-i\left(-\frac{1}{2}\right) \int_{0}^{t}\Delta(t)[M_{\sigma_z}(t)] dt\right],
\end{split}\end{equation} 
and
\begin{equation} \begin{split}
\ket{\varphi^{(1)}(t)}=&\begin{bmatrix}
C_{0+}(0)       & C_{1+}(0)  &   \dots & C_{0-}(0)  & C_{1-}(0)  &  \dots 
\end{bmatrix}
\exp\left \{-i\left (-\frac{1}{2}\right) \int_{0}^{t}\Delta(t)[M_{\sigma_z}(t)] dt\right \}\\
&\times\begin{bmatrix}
\ket{\tilde{E}^{(0)}_{0+}(t)} & e^{-i \omega t}\ket{\tilde{E}^{(0)}_{1+}(t)} &  \dots & \ket{\tilde{E}^{(0)}_{0-}(t)} & e^{-i \omega t}\ket{\tilde{E}^{(0)}_{1-}(t)} \dots 
\end{bmatrix}^T,
\end{split}\end{equation} 
where for generality we can consider the parameter $\Delta$ to be time--dependent ($\Delta=\Delta(t)$).
\end{widetext}

\section{100- and 400-photon cat states \\and the speed of amplification}

%

\begin{figure}
\includegraphics[width=0.5\textwidth]{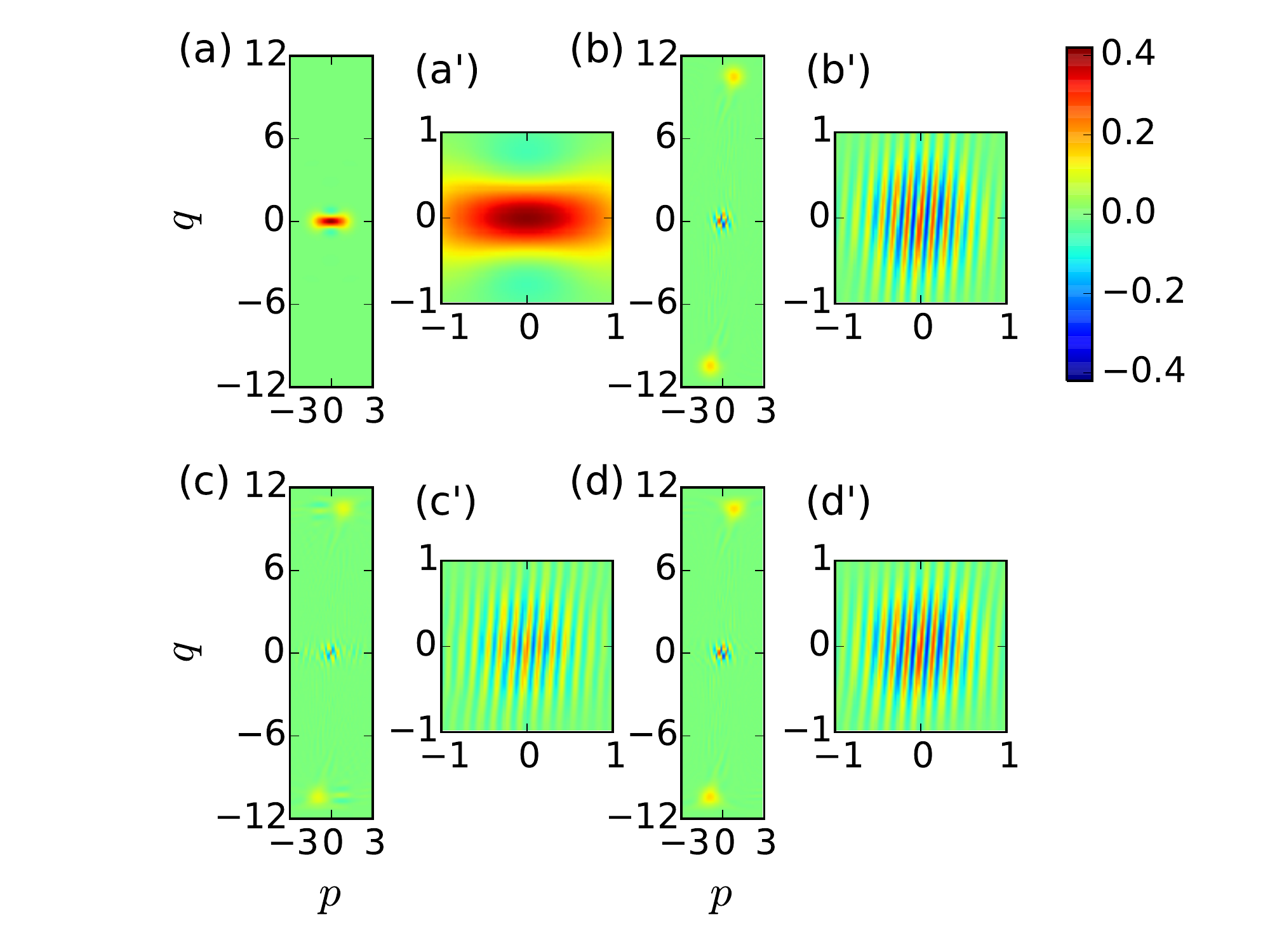}
\caption{\label{fig:400photoncats}
(color online) Cat-state amplification and rephasing in the Wigner representation of the oscillator state projected onto the qubit state 
$\ket{+}_x+\ket{-}_x$. 
The parameters are: $\Delta=0.1\omega$; $g(0)=0.833\omega$, 
and the coupling coefficient is modulated as $g(t)=g(0)\cos\omega t$.
(a) The initial state, which is taken to be the ground state, reasonably resembles the cat state 
$\ket{-\frac{g(0)}{\omega}} + \ket{+\frac{g(0)}{\omega}}$, 
with a fidelity of 
0.99986. 
After the initial state goes through four oscillator periods of the sinusoidally driven cat-state-amplification process, the resulting state is shown in (b), which is the cat state 
$\ket{-\frac{\tilde{g}(8\pi/\omega)}{\omega}} + \ket{+\frac{\tilde{g}(8\pi/\omega)}{\omega}}$, 
with 
$|\tilde{g}/\omega|=10.5$ 
and a fidelity of 
0.988. 
We now let the state in (b) freely evolve for ten oscillator periods, and the resulting state of the oscillator is shown in (c).
The fidelity between the state of (c) and 
the state $\ket{-\frac{\tilde{g}(8\pi/\omega)}{\omega}} + \ket{+\frac{\tilde{g}(8\pi/\omega)}{\omega}}$ 
is only 
0.946, 
since the cat state is dephased by the nonzero $\Delta$.
However, if we insert one $\hat{\sigma}_x$-$\pi$ pulse in the middle 
of the ten oscillator periods, the evolved state, shown in (d), has a 
0.980 
fidelity with the state in (b), analogous to a Hahn spin-echo rephasing effect.
Zoom-in of the central part in (a) to (d) is shown in (a') to (d'), respectively.
}

\end{figure}

In this section, we show simulations of 100- and 400-photon cat states and estimate the time required for the protocol. Compared to the protocol of preparing a 100-photon cat state in Ref. [21], our protocol can be much faster, as we describe below. 
Note that 
the size of a quantum superposition in a cat state $S = |\beta_1 - \beta_2|^2$ is determined by its square distance in phase space between 
the two coherent states in the superposition, 
$\ket{\beta_1}$ and $\ket{\beta_2}$ [11,~21].


To prepare a 100-photon cat state, the protocol in Ref.~[21] takes time $\pi/\chi_{qs}$ at least, where $\chi_{qs}$ is the dispersive interaction rate between the qubit and the oscillator and $\chi_{qs}/2\pi=2.4$ MHz in their case. 
In the case of Fig.~2 in the main text, we modulate the coupling coefficient as 
$g(t)=g(0)\cos\omega t$ 
and our protocol takes $2(2\pi/\omega)$ to reach 
$|\tilde{g}/\omega|=5.3$ 
(Fig.~2), 
which corresponds to a $\sim100$-photon cat state as in Ref. [21], 
where $\omega/2\pi$ is in the order of GHz in typical circuit-QED setups~[21]. 
Assuming a 10 GHz oscillator, our protocol can prepare a 100-photon cat state 
three orders of magnitude 
faster than the setup in Ref. [21].

Furthermore, 
in a protocol that lasts four oscillator periods ($4(2\pi/\omega)$), 
$|\tilde{g}/\omega|$ becomes 10.5 [Figs. 1(c) and 1(d)], which corresponds to a 400-photon cat state (Fig.~\ref{fig:400photoncats}). 

Finally, we emphasize that the time required for our amplification protocol can be shortened by using larger $g/\omega$ and that $g/\omega$ can be easily designed larger especially in the circuit QED systems.

\section{Rephasing during amplification \\($g/\omega=0.1$: smaller coupling case)}


\begin{figure}
\vspace{-20mm}
\includegraphics[width=0.5\textwidth]{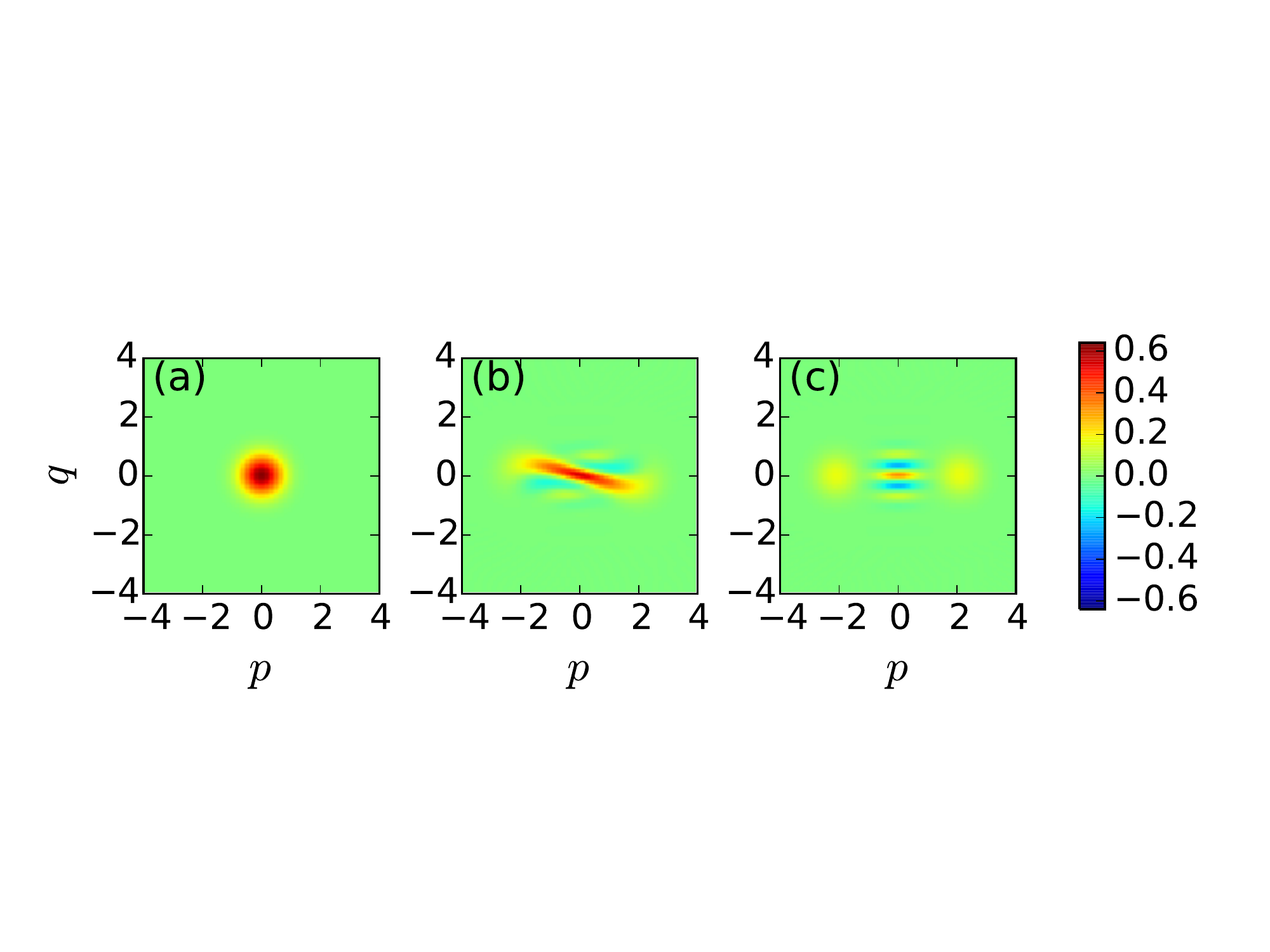}
\vspace{-20mm}
\caption{\label{fig:sigmay}
(color online) Cat-state amplification and rephasing in the Wigner representation of the oscillator state projected onto the qubit state 
$\ket{+}_x+\ket{-}_x$. 
The parameters are: $\Delta=0.1\omega$; $g(0)=0.1\omega$.
(a) The initial state, which is taken to be the ground state, reasonably resembles the cat state 
$\ket{-\frac{g(0)}{\omega}} + \ket{+\frac{g(0)}{\omega}}$, 
with a fidelity of 
1.000. 
(b) After the initial state goes through five oscillator periods of cat-state-amplification process using $\sigma_z$-$\pi$ pulses at every half of the oscillator period, 
the resulting state of the oscillator is shown, 
which is far from the cat state, due to the effect of the finite $\Delta$. 
(c) After the initial state goes through five oscillator periods of cat-state-amplification process using $\sigma_y$-$\pi$ pulses at every half of the oscillator period, 
the resulting state of the oscillator is shown,
which is the cat state, 
$\ket{-\frac{\tilde{g}(10\pi/\omega)}{\omega}} + \ket{+\frac{\tilde{g}(10\pi/\omega)}{\omega}}$, 
with 
$|\tilde{g}/\omega|=2.1$ 
and a fidelity of 
0.999953. 
In (c), by using $\sigma_y$-$\pi$ pulses instead of $\sigma_z$-$\pi$ pulses, cat state amplification and rephasing are simultaneously realized.
}

\end{figure}


In the case of systems with weaker coupling strength, cat state amplification is still achievable, but the time required is longer, and dephasing effect due to $\Delta$ is more apparent. One way to reduce such effect is to employ a smaller $\Delta$. Another way is to partially cancel the dephasing effect while the cat state amplification is ongoing, which can be done by applying $\pi$ pulses to the qubit.

As mentioned in the main text, 
cat states can be amplified by applying $\sigma_z$-$\pi$ pulses on the qubit, and dephasing due to $\Delta$ can be corrected by $\sigma_x$-$\pi$ pulses. Both amplifying a cat state and rephasing can be achieved by $\sigma_y$-$\pi$ pulses.

In this section, we show a simulation of rephasing during amplification, in the case of smaller coupling $g/\omega=0.1$, which is about an order of magnitude smaller than the previous simulations.
The $\sigma_y$-$\pi$ pulses at every half period of the oscillator can cancel the dephasing due to $\Delta$ while amplifying the cat state as shown below.

A simulation of cat state amplification using  $\sigma_z$-$\pi$ (amplifying) and $\sigma_y$-$\pi$ (amplifying and rephasing) pulses, which lasts for five oscillator periods, is shown in Fig.~\ref{fig:sigmay}. The oscillator part of the ground state [Fig.~\ref{fig:sigmay}(a)] 
is close to the vacuum state. 
When $\sigma_z$-$\pi$ (amplifying without rephasing) pulses are used, 
the resulting state contains a finite number of photons but does not have the form of the desired cat state. 
On the other hand, when $\sigma_y$-$\pi$ (amplifying and rephasing) pulses are used, the dephasing due to $\Delta$ is effectively canceled during amplification, and the resulting state is very close to the desired cat state, with a fidelity of 0.999953.


\newpage




\begin{thebibliography}{30}%
\makeatletter
\providecommand \@ifxundefined [1]{%
 \@ifx{#1\undefined}
}%
\providecommand \@ifnum [1]{%
 \ifnum #1\expandafter \@firstoftwo
 \else \expandafter \@secondoftwo
 \fi
}%
\providecommand \@ifx [1]{%
 \ifx #1\expandafter \@firstoftwo
 \else \expandafter \@secondoftwo
 \fi
}%
\providecommand \natexlab [1]{#1}%
\providecommand \enquote  [1]{``#1''}%
\providecommand \bibnamefont  [1]{#1}%
\providecommand \bibfnamefont [1]{#1}%
\providecommand \citenamefont [1]{#1}%
\providecommand \href@noop [0]{\@secondoftwo}%
\providecommand \href [0]{\begingroup \@sanitize@url \@href}%
\providecommand \@href[1]{\@@startlink{#1}\@@href}%
\providecommand \@@href[1]{\endgroup#1\@@endlink}%
\providecommand \@sanitize@url [0]{\catcode `\\12\catcode `\$12\catcode
  `\&12\catcode `\#12\catcode `\^12\catcode `\_12\catcode `\%12\relax}%
\providecommand \@@startlink[1]{}%
\providecommand \@@endlink[0]{}%
\providecommand \url  [0]{\begingroup\@sanitize@url \@url }%
\providecommand \@url [1]{\endgroup\@href {#1}{\urlprefix }}%
\providecommand \urlprefix  [0]{URL }%
\providecommand \Eprint [0]{\href }%
\providecommand \doibase [0]{http://dx.doi.org/}%
\providecommand \selectlanguage [0]{\@gobble}%
\providecommand \bibinfo  [0]{\@secondoftwo}%
\providecommand \bibfield  [0]{\@secondoftwo}%
\providecommand \translation [1]{[#1]}%
\providecommand \BibitemOpen [0]{}%
\providecommand \bibitemStop [0]{}%
\providecommand \bibitemNoStop [0]{.\EOS\space}%
\providecommand \EOS [0]{\spacefactor3000\relax}%
\providecommand \BibitemShut  [1]{\csname bibitem#1\endcsname}%
\let\auto@bib@innerbib\@empty
\bibitem [{\citenamefont {Chiorescu}\ \emph {et~al.}(2004)\citenamefont
  {Chiorescu}, \citenamefont {Bertet}, \citenamefont {Semba}, \citenamefont
  {Nakamura}, \citenamefont {Harmans},\ and\ \citenamefont
  {Mooij}}]{chiorescu2004coherent}%
  \BibitemOpen
  \bibfield  {author} {\bibinfo {author} {\bibfnamefont {I.}~\bibnamefont
  {Chiorescu}}, \bibinfo {author} {\bibfnamefont {P.}~\bibnamefont {Bertet}},
  \bibinfo {author} {\bibfnamefont {K.}~\bibnamefont {Semba}}, \bibinfo
  {author} {\bibfnamefont {Y.}~\bibnamefont {Nakamura}}, \bibinfo {author}
  {\bibfnamefont {C.}~\bibnamefont {Harmans}}, \ and\ \bibinfo {author}
  {\bibfnamefont {J.}~\bibnamefont {Mooij}},\ }\href@noop {} {\bibfield
  {journal} {\bibinfo  {journal} {Nature}\ }\textbf {\bibinfo {volume} {431}},\
  \bibinfo {pages} {159} (\bibinfo {year} {2004})}\BibitemShut {NoStop}%
\bibitem [{\citenamefont {Wallraff}\ \emph {et~al.}(2004)\citenamefont
  {Wallraff}, \citenamefont {Schuster}, \citenamefont {Blais}, \citenamefont
  {Frunzio}, \citenamefont {Huang}, \citenamefont {Majer}, \citenamefont
  {Kumar}, \citenamefont {Girvin},\ and\ \citenamefont
  {Schoelkopf}}]{wallraff2004strong}%
  \BibitemOpen
  \bibfield  {author} {\bibinfo {author} {\bibfnamefont {A.}~\bibnamefont
  {Wallraff}}, \bibinfo {author} {\bibfnamefont {D.~I.}\ \bibnamefont
  {Schuster}}, \bibinfo {author} {\bibfnamefont {A.}~\bibnamefont {Blais}},
  \bibinfo {author} {\bibfnamefont {L.}~\bibnamefont {Frunzio}}, \bibinfo
  {author} {\bibfnamefont {R.-S.}\ \bibnamefont {Huang}}, \bibinfo {author}
  {\bibfnamefont {J.}~\bibnamefont {Majer}}, \bibinfo {author} {\bibfnamefont
  {S.}~\bibnamefont {Kumar}}, \bibinfo {author} {\bibfnamefont {S.~M.}\
  \bibnamefont {Girvin}}, \ and\ \bibinfo {author} {\bibfnamefont {R.~J.}\
  \bibnamefont {Schoelkopf}},\ }\href@noop {} {\bibfield  {journal} {\bibinfo
  {journal} {Nature}\ }\textbf {\bibinfo {volume} {431}},\ \bibinfo {pages}
  {162} (\bibinfo {year} {2004})}\BibitemShut {NoStop}%
\bibitem [{\citenamefont {Devoret}\ \emph {et~al.}(2007)\citenamefont
  {Devoret}, \citenamefont {Girvin},\ and\ \citenamefont
  {Schoelkopf}}]{devoret2007circuit}%
  \BibitemOpen
  \bibfield  {author} {\bibinfo {author} {\bibfnamefont {M.}~\bibnamefont
  {Devoret}}, \bibinfo {author} {\bibfnamefont {S.}~\bibnamefont {Girvin}}, \
  and\ \bibinfo {author} {\bibfnamefont {R.}~\bibnamefont {Schoelkopf}},\
  }\href@noop {} {\bibfield  {journal} {\bibinfo  {journal} {Annalen der
  Physik}\ }\textbf {\bibinfo {volume} {16}},\ \bibinfo {pages} {767} (\bibinfo
  {year} {2007})}\BibitemShut {NoStop}%
\bibitem [{\citenamefont {Niemczyk}\ \emph {et~al.}(2010)\citenamefont
  {Niemczyk}, \citenamefont {Deppe}, \citenamefont {Huebl}, \citenamefont
  {Menzel}, \citenamefont {Hocke}, \citenamefont {Schwarz}, \citenamefont
  {Garcia-Ripoll}, \citenamefont {Zueco}, \citenamefont {H{\"u}mmer},
  \citenamefont {Solano} \emph {et~al.}}]{niemczyk2010circuit}%
  \BibitemOpen
  \bibfield  {author} {\bibinfo {author} {\bibfnamefont {T.}~\bibnamefont
  {Niemczyk}}, \bibinfo {author} {\bibfnamefont {F.}~\bibnamefont {Deppe}},
  \bibinfo {author} {\bibfnamefont {H.}~\bibnamefont {Huebl}}, \bibinfo
  {author} {\bibfnamefont {E.}~\bibnamefont {Menzel}}, \bibinfo {author}
  {\bibfnamefont {F.}~\bibnamefont {Hocke}}, \bibinfo {author} {\bibfnamefont
  {M.}~\bibnamefont {Schwarz}}, \bibinfo {author} {\bibfnamefont
  {J.}~\bibnamefont {Garcia-Ripoll}}, \bibinfo {author} {\bibfnamefont
  {D.}~\bibnamefont {Zueco}}, \bibinfo {author} {\bibfnamefont
  {T.}~\bibnamefont {H{\"u}mmer}}, \bibinfo {author} {\bibfnamefont
  {E.}~\bibnamefont {Solano}},  \emph {et~al.},\ }\href@noop {} {\bibfield
  {journal} {\bibinfo  {journal} {Nature Phys.}\ }\textbf {\bibinfo {volume}
  {6}},\ \bibinfo {pages} {772} (\bibinfo {year} {2010})}\BibitemShut {NoStop}%
\bibitem [{\citenamefont {Forn-D{\'\i}az}\ \emph {et~al.}(2010)\citenamefont
  {Forn-D{\'\i}az}, \citenamefont {Lisenfeld}, \citenamefont {Marcos},
  \citenamefont {Garc{\'\i}a-Ripoll}, \citenamefont {Solano}, \citenamefont
  {Harmans},\ and\ \citenamefont {Mooij}}]{forn2010observation}%
  \BibitemOpen
  \bibfield  {author} {\bibinfo {author} {\bibfnamefont {P.}~\bibnamefont
  {Forn-D{\'\i}az}}, \bibinfo {author} {\bibfnamefont {J.}~\bibnamefont
  {Lisenfeld}}, \bibinfo {author} {\bibfnamefont {D.}~\bibnamefont {Marcos}},
  \bibinfo {author} {\bibfnamefont {J.~J.}\ \bibnamefont {Garc{\'\i}a-Ripoll}},
  \bibinfo {author} {\bibfnamefont {E.}~\bibnamefont {Solano}}, \bibinfo
  {author} {\bibfnamefont {C.~J.~P.~M.}~\bibnamefont {Harmans}}, \ and\ \bibinfo
  {author} {\bibfnamefont {J.~E.}~\bibnamefont {Mooij}},\ }\href@noop {}
  {\bibfield  {journal} {\bibinfo  {journal} {Phys. Rev. Lett.}\ }\textbf
  {\bibinfo {volume} {105}},\ \bibinfo {pages} {237001} (\bibinfo {year}
  {2010})}\BibitemShut {NoStop}%
\bibitem [{\citenamefont {Yoshihara}\ \emph
  {et~al.}(2017{\natexlab{a}})\citenamefont {Yoshihara}, \citenamefont {Fuse},
  \citenamefont {Ashhab}, \citenamefont {Kakuyanagi}, \citenamefont {Saito},\
  and\ \citenamefont {Semba}}]{yoshihara2017superconducting}%
  \BibitemOpen
  \bibfield  {author} {\bibinfo {author} {\bibfnamefont {F.}~\bibnamefont
  {Yoshihara}}, \bibinfo {author} {\bibfnamefont {T.}~\bibnamefont {Fuse}},
  \bibinfo {author} {\bibfnamefont {S.}~\bibnamefont {Ashhab}}, \bibinfo
  {author} {\bibfnamefont {K.}~\bibnamefont {Kakuyanagi}}, \bibinfo {author}
  {\bibfnamefont {S.}~\bibnamefont {Saito}}, \ and\ \bibinfo {author}
  {\bibfnamefont {K.}~\bibnamefont {Semba}},\ }\href@noop {} {\bibfield
  {journal} {\bibinfo  {journal} {Nature Phys.}\ }\textbf {\bibinfo {volume}
  {13}},\ \bibinfo {pages} {44} (\bibinfo {year}
  {2017}{\natexlab{a}})}\BibitemShut {NoStop}%
\bibitem [{\citenamefont {Yoshihara}\ \emph
  {et~al.}(2017{\natexlab{b}})\citenamefont {Yoshihara}, \citenamefont {Fuse},
  \citenamefont {Ashhab}, \citenamefont {Kakuyanagi}, \citenamefont {Saito},\
  and\ \citenamefont {Semba}}]{yoshihara2017characteristic}%
  \BibitemOpen
  \bibfield  {author} {\bibinfo {author} {\bibfnamefont {F.}~\bibnamefont
  {Yoshihara}}, \bibinfo {author} {\bibfnamefont {T.}~\bibnamefont {Fuse}},
  \bibinfo {author} {\bibfnamefont {S.}~\bibnamefont {Ashhab}}, \bibinfo
  {author} {\bibfnamefont {K.}~\bibnamefont {Kakuyanagi}}, \bibinfo {author}
  {\bibfnamefont {S.}~\bibnamefont {Saito}}, \ and\ \bibinfo {author}
  {\bibfnamefont {K.}~\bibnamefont {Semba}},\ }\href@noop {} {\bibfield
  {journal} {\bibinfo  {journal} {Phys. Rev. A}\ }\textbf {\bibinfo {volume}
  {95}},\ \bibinfo {pages} {053824} (\bibinfo {year}
  {2017}{\natexlab{b}})}\BibitemShut {NoStop}%
\bibitem [{\citenamefont {Didier}\ \emph {et~al.}(2015)\citenamefont {Didier},
  \citenamefont {Bourassa},\ and\ \citenamefont {Blais}}]{didier2015fast}%
  \BibitemOpen
  \bibfield  {author} {\bibinfo {author} {\bibfnamefont {N.}~\bibnamefont
  {Didier}}, \bibinfo {author} {\bibfnamefont {J.}~\bibnamefont {Bourassa}}, \
  and\ \bibinfo {author} {\bibfnamefont {A.}~\bibnamefont {Blais}},\
  }\href@noop {} {\bibfield  {journal} {\bibinfo  {journal} {Phys. Rev. Lett.}\
  }\textbf {\bibinfo {volume} {115}},\ \bibinfo {pages} {203601} (\bibinfo
  {year} {2015})}\BibitemShut {NoStop}%
\bibitem [{\citenamefont {Yin}\ \emph {et~al.}(2012)\citenamefont {Yin},
  \citenamefont {Wang}, \citenamefont {Mariantoni}, \citenamefont {Bialczak},
  \citenamefont {Barends}, \citenamefont {Chen}, \citenamefont {Lenander},
  \citenamefont {Lucero}, \citenamefont {Neeley}, \citenamefont {O'Connell}
  \emph {et~al.}}]{yin2012dynamic}%
  \BibitemOpen
  \bibfield  {author} {\bibinfo {author} {\bibfnamefont {Y.}~\bibnamefont
  {Yin}}, \bibinfo {author} {\bibfnamefont {H.}~\bibnamefont {Wang}}, \bibinfo
  {author} {\bibfnamefont {M.}~\bibnamefont {Mariantoni}}, \bibinfo {author}
  {\bibfnamefont {R.~C.}\ \bibnamefont {Bialczak}}, \bibinfo {author}
  {\bibfnamefont {R.}~\bibnamefont {Barends}}, \bibinfo {author} {\bibfnamefont
  {Y.}~\bibnamefont {Chen}}, \bibinfo {author} {\bibfnamefont {M.}~\bibnamefont
  {Lenander}}, \bibinfo {author} {\bibfnamefont {E.}~\bibnamefont {Lucero}},
  \bibinfo {author} {\bibfnamefont {M.}~\bibnamefont {Neeley}}, \bibinfo
  {author} {\bibfnamefont {A.}~\bibnamefont {O'Connell}},  \emph {et~al.},\
  }\href@noop {} {\bibfield  {journal} {\bibinfo  {journal} {Phys. Rev. A}\
  }\textbf {\bibinfo {volume} {85}},\ \bibinfo {pages} {023826} (\bibinfo
  {year} {2012})}\BibitemShut {NoStop}%
\bibitem [{\citenamefont {Gu}\ \emph {et~al.}(2017)\citenamefont {Gu},
  \citenamefont {Kockum}, \citenamefont {Miranowicz}, \citenamefont {xi~Liu},\
  and\ \citenamefont {Nori}}]{Nori2017review}%
  \BibitemOpen
  \bibfield  {author} {\bibinfo {author} {\bibfnamefont {X.}~\bibnamefont
  {Gu}}, \bibinfo {author} {\bibfnamefont {A.~F.}\ \bibnamefont {Kockum}},
  \bibinfo {author} {\bibfnamefont {A.}~\bibnamefont {Miranowicz}}, \bibinfo
  {author} {\bibfnamefont {Y.}~\bibnamefont {xi~Liu}}, \ and\ \bibinfo {author}
  {\bibfnamefont {F.}~\bibnamefont {Nori}},\ }\href@noop {} {\bibfield
  {journal} {\bibinfo  {journal} {Physics Reports}\ }\textbf {\bibinfo {volume}
  {718-719}},\ \bibinfo {pages} {1 } (\bibinfo {year} {2017})}\BibitemShut
  {NoStop}%
\bibitem [{\citenamefont {Deleglise}\ \emph {et~al.}(2008)\citenamefont
  {Deleglise}, \citenamefont {Dotsenko}, \citenamefont {Sayrin}, \citenamefont
  {Bernu}, \citenamefont {Brune}, \citenamefont {Raimond},\ and\ \citenamefont
  {Haroche}}]{Haroche2008reconstruction}%
  \BibitemOpen
  \bibfield  {author} {\bibinfo {author} {\bibfnamefont {S.}~\bibnamefont
  {Deleglise}}, \bibinfo {author} {\bibfnamefont {I.}~\bibnamefont {Dotsenko}},
  \bibinfo {author} {\bibfnamefont {C.}~\bibnamefont {Sayrin}}, \bibinfo
  {author} {\bibfnamefont {J.}~\bibnamefont {Bernu}}, \bibinfo {author}
  {\bibfnamefont {M.}~\bibnamefont {Brune}}, \bibinfo {author} {\bibfnamefont
  {J.-M.}\ \bibnamefont {Raimond}}, \ and\ \bibinfo {author} {\bibfnamefont
  {S.}~\bibnamefont {Haroche}},\ }\href@noop {} {\bibfield  {journal} {\bibinfo
   {journal} {Nature}\ }\textbf {\bibinfo {volume} {455}},\ \bibinfo {pages}
  {510} (\bibinfo {year} {2008})}\BibitemShut {NoStop}%
\bibitem [{\citenamefont {Haroche}\ and\ \citenamefont
  {Raimond}(2013)}]{Haroche}%
  \BibitemOpen
  \bibfield  {author} {\bibinfo {author} {\bibfnamefont {S.}~\bibnamefont
  {Haroche}}\ and\ \bibinfo {author} {\bibfnamefont {J.-M.}\ \bibnamefont
  {Raimond}},\ }\href@noop {} {\emph {\bibinfo {title} {Exploring the Quantum:
  Atoms, Cavities, and Photons.}}}\ (\bibinfo  {publisher} {Oxford Graduate Texts},\ \bibinfo
  {year} {2013})\ Chap.~\bibinfo {chapter} {7.}\BibitemShut {Stop}%
\bibitem [{\citenamefont {Irish}(2007)}]{Irish2007gRWA}%
  \BibitemOpen
  \bibfield  {author} {\bibinfo {author} {\bibfnamefont {E.~K.}\ \bibnamefont
  {Irish}},\ }\href@noop {} {\bibfield  {journal} {\bibinfo  {journal} {Phys.
  Rev. Lett.}\ }\textbf {\bibinfo {volume} {99}},\ \bibinfo {pages} {173601}
  (\bibinfo {year} {2007})}\BibitemShut {NoStop}%
\bibitem [{\citenamefont {Braak}(2011)}]{Braak2011}%
  \BibitemOpen
  \bibfield  {author} {\bibinfo {author} {\bibfnamefont {D.}~\bibnamefont
  {Braak}},\ }\href@noop {} {\bibfield  {journal} {\bibinfo  {journal} {Phys.
  Rev. Lett.}\ }\textbf {\bibinfo {volume} {107}},\ \bibinfo {pages} {100401}
  (\bibinfo {year} {2011})}\BibitemShut {NoStop}%
\bibitem [{\citenamefont {Ashhab}\ and\ \citenamefont
  {Nori}(2010)}]{ashhab2010qubit}%
  \BibitemOpen
  \bibfield  {author} {\bibinfo {author} {\bibfnamefont {S.}~\bibnamefont
  {Ashhab}}\ and\ \bibinfo {author} {\bibfnamefont {F.}~\bibnamefont {Nori}},\
  }\href@noop {} {\bibfield  {journal} {\bibinfo  {journal} {Phys. Rev. A}\
  }\textbf {\bibinfo {volume} {81}},\ \bibinfo {pages} {042311} (\bibinfo
  {year} {2010})}\BibitemShut {NoStop}%
\bibitem [{\citenamefont {Casanova}\ \emph {et~al.}(2010)\citenamefont
  {Casanova}, \citenamefont {Romero}, \citenamefont {Lizuain}, \citenamefont
  {Garc{\'\i}a-Ripoll},\ and\ \citenamefont {Solano}}]{casanova2010deep}%
  \BibitemOpen
  \bibfield  {author} {\bibinfo {author} {\bibfnamefont {J.}~\bibnamefont
  {Casanova}}, \bibinfo {author} {\bibfnamefont {G.}~\bibnamefont {Romero}},
  \bibinfo {author} {\bibfnamefont {I.}~\bibnamefont {Lizuain}}, \bibinfo
  {author} {\bibfnamefont {J.~J.}~\bibnamefont {Garc{\'\i}a-Ripoll}}, \ and\
  \bibinfo {author} {\bibfnamefont {E.}~\bibnamefont {Solano}},\ }\href@noop {}
  {\bibfield  {journal} {\bibinfo  {journal} {Phys. Rev. Lett.}\ }\textbf
  {\bibinfo {volume} {105}},\ \bibinfo {pages} {263603} (\bibinfo {year}
  {2010})}\BibitemShut {NoStop}%
\bibitem [{\citenamefont {Schuetz}\ \emph {et~al.}(2017)\citenamefont
  {Schuetz}, \citenamefont {Giedke}, \citenamefont {Vandersypen},\ and\
  \citenamefont {Cirac}}]{Schuetz2017}%
  \BibitemOpen
  \bibfield  {author} {\bibinfo {author} {\bibfnamefont {M.~J.~A.}\
  \bibnamefont {Schuetz}}, \bibinfo {author} {\bibfnamefont {G.}~\bibnamefont
  {Giedke}}, \bibinfo {author} {\bibfnamefont {L.~M.~K.}\ \bibnamefont
  {Vandersypen}}, \ and\ \bibinfo {author} {\bibfnamefont {J.~I.}\ \bibnamefont
  {Cirac}},\ }\href@noop {} {\bibfield  {journal} {\bibinfo  {journal} {Phys.
  Rev. A}\ }\textbf {\bibinfo {volume} {95}},\ \bibinfo {pages} {052335}
  (\bibinfo {year} {2017})}\BibitemShut {NoStop}%
\bibitem [{\citenamefont {Wang}\ \emph {et~al.}(2009)\citenamefont {Wang},
  \citenamefont {Kemp},\ and\ \citenamefont {Semba}}]{wang2009coupling}%
  \BibitemOpen
  \bibfield  {author} {\bibinfo {author} {\bibfnamefont {Y.-D.}\ \bibnamefont
  {Wang}}, \bibinfo {author} {\bibfnamefont {A.}~\bibnamefont {Kemp}}, \ and\
  \bibinfo {author} {\bibfnamefont {K.}~\bibnamefont {Semba}},\ }\href@noop {}
  {\bibfield  {journal} {\bibinfo  {journal} {Phys. Rev. B}\ }\textbf {\bibinfo
  {volume} {79}},\ \bibinfo {pages} {024502} (\bibinfo {year}
  {2009})}\BibitemShut {NoStop}%
\bibitem [{\citenamefont {Leroux}\ \emph {et~al.}(2018)\citenamefont {Leroux},
  \citenamefont {Govia},\ and\ \citenamefont {Clerk}}]{Clerk2018}%
  \BibitemOpen
  \bibfield  {author} {\bibinfo {author} {\bibfnamefont {C.}~\bibnamefont
  {Leroux}}, \bibinfo {author} {\bibfnamefont {L.~C.~G.}\ \bibnamefont
  {Govia}}, \ and\ \bibinfo {author} {\bibfnamefont {A.~A.}\ \bibnamefont
  {Clerk}},\ }\href@noop {} {\bibfield  {journal} {\bibinfo  {journal} {Phys.
  Rev. Lett.}\ }\textbf {\bibinfo {volume} {120}},\ \bibinfo {pages} {093602}
  (\bibinfo {year} {2018})}\BibitemShut {NoStop}%
\bibitem [{\citenamefont {Hofheinz}\ \emph {et~al.}(2009)\citenamefont
  {Hofheinz}, \citenamefont {Wang}, \citenamefont {Ansmann}, \citenamefont
  {Bialczak}, \citenamefont {Lucero}, \citenamefont {Neeley}, \citenamefont
  {O'connell}, \citenamefont {Sank}, \citenamefont {Wenner}, \citenamefont
  {Martinis} \emph {et~al.}}]{hofheinz2009synthesizing}%
  \BibitemOpen
  \bibfield  {author} {\bibinfo {author} {\bibfnamefont {M.}~\bibnamefont
  {Hofheinz}}, \bibinfo {author} {\bibfnamefont {H.}~\bibnamefont {Wang}},
  \bibinfo {author} {\bibfnamefont {M.}~\bibnamefont {Ansmann}}, \bibinfo
  {author} {\bibfnamefont {R.~C.}\ \bibnamefont {Bialczak}}, \bibinfo {author}
  {\bibfnamefont {E.}~\bibnamefont {Lucero}}, \bibinfo {author} {\bibfnamefont
  {M.}~\bibnamefont {Neeley}}, \bibinfo {author} {\bibfnamefont
  {A.}~\bibnamefont {O'connell}}, \bibinfo {author} {\bibfnamefont
  {D.}~\bibnamefont {Sank}}, \bibinfo {author} {\bibfnamefont {J.}~\bibnamefont
  {Wenner}}, \bibinfo {author} {\bibfnamefont {J.~M.}\ \bibnamefont
  {Martinis}},  \emph {et~al.},\ }\href@noop {} {\bibfield  {journal} {\bibinfo
   {journal} {Nature}\ }\textbf {\bibinfo {volume} {459}},\ \bibinfo {pages}
  {546} (\bibinfo {year} {2009})}\BibitemShut {NoStop}%
\bibitem [{\citenamefont {Vlastakis}\ \emph {et~al.}(2013)\citenamefont
  {Vlastakis}, \citenamefont {Kirchmair}, \citenamefont {Leghtas},
  \citenamefont {Nigg}, \citenamefont {Frunzio}, \citenamefont {Girvin},
  \citenamefont {Mirrahimi}, \citenamefont {Devoret},\ and\ \citenamefont
  {Schoelkopf}}]{vlastakis2013deterministically}%
  \BibitemOpen
  \bibfield  {author} {\bibinfo {author} {\bibfnamefont {B.}~\bibnamefont
  {Vlastakis}}, \bibinfo {author} {\bibfnamefont {G.}~\bibnamefont
  {Kirchmair}}, \bibinfo {author} {\bibfnamefont {Z.}~\bibnamefont {Leghtas}},
  \bibinfo {author} {\bibfnamefont {S.~E.}\ \bibnamefont {Nigg}}, \bibinfo
  {author} {\bibfnamefont {L.}~\bibnamefont {Frunzio}}, \bibinfo {author}
  {\bibfnamefont {S.~M.}\ \bibnamefont {Girvin}}, \bibinfo {author}
  {\bibfnamefont {M.}~\bibnamefont {Mirrahimi}}, \bibinfo {author}
  {\bibfnamefont {M.~H.}\ \bibnamefont {Devoret}}, \ and\ \bibinfo {author}
  {\bibfnamefont {R.~J.}\ \bibnamefont {Schoelkopf}},\ }\href@noop {}
  {\bibfield  {journal} {\bibinfo  {journal} {Science}\ }\textbf {\bibinfo
  {volume} {342}},\ \bibinfo {pages} {607} (\bibinfo {year}
  {2013})}\BibitemShut {NoStop}%
\bibitem [{\citenamefont {Facon}\ \emph {et~al.}(2016)\citenamefont {Facon},
  \citenamefont {Dietsche}, \citenamefont {Grosso}, \citenamefont {Haroche},
  \citenamefont {Raimond}, \citenamefont {Brune},\ and\ \citenamefont
  {Gleyzes}}]{Haroche2016sensing}%
  \BibitemOpen
  \bibfield  {author} {\bibinfo {author} {\bibfnamefont {A.}~\bibnamefont
  {Facon}}, \bibinfo {author} {\bibfnamefont {E.-K.}\ \bibnamefont {Dietsche}},
  \bibinfo {author} {\bibfnamefont {D.}~\bibnamefont {Grosso}}, \bibinfo
  {author} {\bibfnamefont {S.}~\bibnamefont {Haroche}}, \bibinfo {author}
  {\bibfnamefont {J.-M.}\ \bibnamefont {Raimond}}, \bibinfo {author}
  {\bibfnamefont {M.}~\bibnamefont {Brune}}, \ and\ \bibinfo {author}
  {\bibfnamefont {S.}~\bibnamefont {Gleyzes}},\ }\href@noop {} {\bibfield
  {journal} {\bibinfo  {journal} {Nature}\ }\textbf {\bibinfo {volume} {535}},\
  \bibinfo {pages} {262} (\bibinfo {year} {2016})}\BibitemShut {NoStop}%
\bibitem [{\citenamefont {Braak}\ \emph {et~al.}(2016)\citenamefont {Braak},
  \citenamefont {Chen}, \citenamefont {Batchelor},\ and\ \citenamefont
  {Solano}}]{RabiModel}%
  \BibitemOpen
  \bibfield  {author} {\bibinfo {author} {\bibfnamefont {D.}~\bibnamefont
  {Braak}}, \bibinfo {author} {\bibfnamefont {Q.-H.}\ \bibnamefont {Chen}},
  \bibinfo {author} {\bibfnamefont {M.~T.}\ \bibnamefont {Batchelor}}, \ and\
  \bibinfo {author} {\bibfnamefont {E.}~\bibnamefont {Solano}},\ }\href@noop {}
  {\bibfield  {journal} {\bibinfo  {journal} {Journal of Physics A:
  Mathematical and Theoretical}\ }\textbf {\bibinfo {volume} {49}},\ \bibinfo
  {pages} {300301} (\bibinfo {year} {2016})}\BibitemShut {NoStop}%
\bibitem [{\citenamefont {Schr\"{o}dinger}(1926)}]{Schrodinger1926}%
  \BibitemOpen
  \bibfield  {author} {\bibinfo {author} {\bibfnamefont {E.}~\bibnamefont
  {Schr\"{o}dinger}},\ }\href@noop {} {\bibfield  {journal} {\bibinfo
  {journal} {Naturwissenschaften}\ }\textbf {\bibinfo {volume} {14}},\ \bibinfo
  {pages} {664} (\bibinfo {year} {1926})}\BibitemShut {NoStop}%
\bibitem [{\citenamefont {Kodach}\ \emph {et~al.}(2010)\citenamefont {Kodach},
  \citenamefont {Kalkman}, \citenamefont {Faber},\ and\ \citenamefont {van
  Leeuwen}}]{Kodach:10}%
  \BibitemOpen
  \bibfield  {author} {\bibinfo {author} {\bibfnamefont {V.~M.}\ \bibnamefont
  {Kodach}}, \bibinfo {author} {\bibfnamefont {J.}~\bibnamefont {Kalkman}},
  \bibinfo {author} {\bibfnamefont {D.~J.}\ \bibnamefont {Faber}}, \ and\
  \bibinfo {author} {\bibfnamefont {T.~G.}\ \bibnamefont {van Leeuwen}},\
  }\href@noop {} {\bibfield  {journal} {\bibinfo  {journal} {Biomed. Opt.
  Express}\ }\textbf {\bibinfo {volume} {1}},\ \bibinfo {pages} {176} (\bibinfo
  {year} {2010})}\BibitemShut {NoStop}%
\bibitem [{\citenamefont {Viola}\ \emph {et~al.}(1999)\citenamefont {Viola},
  \citenamefont {Knill},\ and\ \citenamefont {Lloyd}}]{viola1999viola}%
  \BibitemOpen
  \bibfield  {author} {\bibinfo {author} {\bibfnamefont {L.}~\bibnamefont
  {Viola}}, \bibinfo {author} {\bibfnamefont {E.}~\bibnamefont {Knill}}, \ and\
  \bibinfo {author} {\bibfnamefont {S.}~\bibnamefont {Lloyd}},\ }\href@noop {}
  {\bibfield  {journal} {\bibinfo  {journal} {Phys. Rev. Lett.}\ }\textbf
  {\bibinfo {volume} {82}},\ \bibinfo {pages} {2417} (\bibinfo {year}
  {1999})}\BibitemShut {NoStop}%
\bibitem [{\citenamefont {de~Lange}\ \emph {et~al.}(2010)\citenamefont
  {de~Lange}, \citenamefont {Wang}, \citenamefont {Rist{\`e}}, \citenamefont
  {Dobrovitski},\ and\ \citenamefont {Hanson}}]{de2010g}%
  \BibitemOpen
  \bibfield  {author} {\bibinfo {author} {\bibfnamefont {G.}~\bibnamefont
  {de~Lange}}, \bibinfo {author} {\bibfnamefont {Z.~H.}\ \bibnamefont {Wang}},
  \bibinfo {author} {\bibfnamefont {D.}~\bibnamefont {Rist{\`e}}}, \bibinfo
  {author} {\bibfnamefont {V.~V.}\ \bibnamefont {Dobrovitski}}, \ and\ \bibinfo
  {author} {\bibfnamefont {R.}~\bibnamefont {Hanson}},\ }\href@noop {}
  {\bibfield  {journal} {\bibinfo  {journal} {Science}\ }\textbf {\bibinfo
  {volume} {330}},\ \bibinfo {pages} {60} (\bibinfo {year} {2010})}\BibitemShut
  {NoStop}%
\bibitem [{\citenamefont {Hahn}(1950)}]{Hahn50PR}%
  \BibitemOpen
  \bibfield  {author} {\bibinfo {author} {\bibfnamefont {E.~L.}\ \bibnamefont
  {Hahn}},\ }\href@noop {} {\bibfield  {journal} {\bibinfo  {journal} {Phys.\
  Rev.}\ }\textbf {\bibinfo {volume} {80}},\ \bibinfo {pages} {580} (\bibinfo
  {year} {1950})}\BibitemShut {NoStop}%
\bibitem [{\citenamefont {Johansson}\ \emph {et~al.}(2012)\citenamefont
  {Johansson}, \citenamefont {Nation},\ and\ \citenamefont {Nori}}]{qutip}%
  \BibitemOpen
  \bibfield  {author} {\bibinfo {author} {\bibfnamefont {J.}~\bibnamefont
  {Johansson}}, \bibinfo {author} {\bibfnamefont {P.}~\bibnamefont {Nation}}, \
  and\ \bibinfo {author} {\bibfnamefont {F.}~\bibnamefont {Nori}},\ }\href@noop
  {} {\bibfield  {journal} {\bibinfo  {journal} {Computer Physics
  Communications}\ }\textbf {\bibinfo {volume} {183}},\ \bibinfo {pages} {1760
  } (\bibinfo {year} {2012})}\BibitemShut {NoStop}%
\end{thebibliography}

%


\end{document}